\documentclass[fleqn,usenatbib]{mnras}

\usepackage{newtxtext,newtxmath}
\usepackage[T1]{fontenc}
\usepackage{ae,aecompl}
\usepackage{graphicx}	
\usepackage{amsmath}	
\usepackage{amssymb}	

\pdfoutput=1

\title[Modelling Molecular Cloud Dust Polarization]{Modelling Dust Polarization Observations of Molecular Clouds through MHD Simulations}

\author[King et al.]{
Patrick K. King,$^{1,2}$\thanks{E-mail: pkk4hu@virginia.edu}
Laura M. Fissel,$^{3}$
Che-Yu Chen,$^{1}$
and Zhi-Yun Li$^{1}$
\\
$^{1}$Department of Astronomy, University of Virginia, Charlottesville, VA, 22904\\
$^{2}$Harvard-Smithsonian Center for Astrophysics, Cambridge, MA, 02138\\
$^{3}$National Radio Astronomy Observatory, Charlottesville, VA, 22904
}

\date{Accepted XXX. Received YYY; in original form ZZZ}

\pubyear{2017}

\begin{document}
\label{firstpage}
\pagerange{\pageref{firstpage}--\pageref{lastpage}}
\maketitle

\begin{abstract} 
The BLASTPol observations of Vela C have provided the most detailed characterization of the polarization fraction $p$ and dispersion in polarization angles $S$ for a molecular cloud. We compare the observed distributions of $p$ and $S$ with those obtained in synthetic observations of simulations of molecular clouds, assuming homogeneous grain alignment. We find that the orientation of the mean magnetic field relative to the observer has a significant effect on the $p$ and $S$ distributions. These distributions for Vela C are most consistent with synthetic observations where the mean magnetic field is close to the line-of-sight. Our results point to apparent magnetic disorder in the Vela C molecular cloud, although it can be due to either an inclination effect (i.e., observing close to the mean field direction) or significant field tangling from strong turbulence/low magnetization. The joint correlations of $p$ with column density and of $S$ with column density for the synthetic observations generally agree poorly with the Vela C joint correlations, suggesting that understanding these correlations require a more sophisticated treatment of grain alignment physics. 
\end{abstract}


\section{Introduction} \label{section:intro}

Stars are typically formed in molecular clouds (MCs), as a result of runaway collapse in which gas self-gravity overwhelms thermal and magnetic pressure support \citep{S1,M1}. This process is sensitive to the local magnetic field conditions, including magnetic field strength and the relative organization of the magnetic field (herein, collectively referred to as magnetic structure) within the MC \citep{M1,C1}. Magnetic fields provide direct resistance against gravitational collapse; magnetic tension forces tend to constrain gas motions; and magnetized media support additional wave interactions (Alfv\'{e}n and magnetosonic waves.) These effects vary in importance across many scales of hierarchical structure, and if sufficiently strong can even halt gravitational collapse and prevent star formation \citep{MS1,STR,MS2}. In short, a comprehensive understanding of star formation requires a detailed understanding of magnetic structure in MCs. 

On the other hand, magnetic structure is difficult to ascertain, being relatively inaccessible to observations. Through measurement of polarized dust emission, far-IR and submillimeter polarimetry has emerged as a most promising technique for accessing MC magnetic structure \citep{HDN,H,N,DDDSH,H4,MMFC}. Dust grains in MCs are thought to align, on average, perpendicular to the local magnetic field \citep{DG,L,ALV}, and therefore provide a  measure of line-of-sight averaged magnetic structure, in particular the orientation of the magnetic field projected onto the plane-of-sky. Maps of polarized dust emission provide crucial means to test theoretical expectations of magnetic structure \citep{CH,HULL,HBL,Z,COX,CHI,PXXXIII}. However, use of this information has been limited in the past, as each individual polarization pseudovector is difficult to interpret alone, and is necessarily limited by both projection and uncertainties in the physics of grain alignment. 

The magnetic structure of MCs can be studied using numerical simulations, which provide a fully three-dimensional picture of both the gas and magnetic field structure \citep{OSG1,LNMH,NL,LMK}. Using these simulations we can compute synthetic observations under simplifying assumptions about grain alignment physics, which can be directly compared with observational data. Examining both the detailed three-dimensional magnetic structure and the resulting two-dimensional polarimetric observables can help determine specifically dynamical effects (arising purely from gas and magnetic structure) and help disentangle them from effects that might arise exclusively due to grain alignment physics. These numerical efforts have considerable statistical power due to the high theoretically achievable spatial resolution, but nevertheless suffer from a fundamental limitation: without observations that have comparable statistical power, it is unclear whether these models are reasonable models of real MCs. 

Until recently high-resolution sub-mm maps were only available from ground-based polarimeters, which were limited to observing either extremely bright evolved molecular clouds, or small areas within clouds because of atmospheric transmission and loading. Recently the Planck Satellite has produced all-sky polarization maps at 850 $\mu$m \citep{PXIX}, which has enabled detailed studies of magnetic field morphology for nearby low-mass clouds at 10' resolution \citep{PXXXV}. In addition the Balloon-borne Large Aperture Submillimeter Telescope for Polarimetry \citep{GBP}, has produced a similarly detailed map for the more distant ($\sim$ 700 pc) early-stage giant molecular cloud Vela C. With over a thousand independent polarization pseudovector measurements and 0.5 parsec resolution, these observations represent a crucial advance towards achieving statistical power parity with numerical simulations. Future flights with the next generation BLAST-TNG instrument promise observations of more star-forming MC targets with even higher sensitivity.

We present the first detailed statistical comparison of the BLASTPol observations with synthetic polarimetric observations of numerical simulations of star-forming regions, which were conducted using the \textsc{Athena} code \citep{ATHENA}. We used a colliding-flow, oblique MHD shock set-up that was used in \citet{CKL} and \citet{CO1,CO2}. This geometry is motivated by the observation that it is typically in regions with large-scale convergent flows that the gas becomes significantly compressed and results in gravitational instability to collapse \citep{MLK,BKMV}. We use the post-shock region of these simulations as idealized models of a subset of a star-forming MC. We characterize the synthetic and BLASTPol observational datasets statistically, determining both the probability distributions of individual observables and joint correlations between observables. 

Our paper is organized as follows. In Section \ref{section:methods} we describe the BLASTPol observations (Section \ref{section:BP}) and the numerical simulations (Section \ref{section:sims}) used in our comparison, as well as describe the methods used to produce the synthetic polarimetric observations (Section \ref{section:synthpol}) and the statistical techniques we used to analyse both datasets (Section \ref{section:stats}). Next we discuss the probability distributions of the polarization fraction in Section \ref{section:p}. We discuss the probability distributions of the dispersion in polarization angles in Section \ref{section:S}. Next we consider the joint correlation between the polarization fraction and angle dispersion in Section \ref{section:Sp} We consider joint correlations between the column density and the polarimetric observables in Section \ref{section:N}. We discuss the effect of intermediate inclination of the line-of-sight in Section \ref{section:inc}. Finally we conclude and summarize in Section \ref{section:conc}.

\section{Methods} \label{section:methods}

\subsection{BLASTPol Observations of Vela C} \label{section:BP}

BLASTPol is a high altitude balloon-borne polarimeter, which maps the sky simultaneously in three wide frequency bands ($\Delta f/f\,\approx\,30\%$)  centred at 250, 350, and 500 $\mu$m \citep{GBP}.  These frequency bands span the spectral peak of cold (10 to 20\,K) dust, and since the instrument is not limited by atmospheric loading it has both higher sensitivity and can recover larger scale structures than ground based polarimeters.

On 26 December 2012 BLASTPol launched from the NASA Long Duration Balloon Facility near McMurdo Station and attained an average altitude of $\sim$38.5\,km.  The primary science target of the BLASTPol 2012 flight was the nearby Vela\,C giant molecular cloud (d$\,\sim\,$700\,$\pm$\,200\,pc; \citealt{LLNSM}).  Vela\,C is a massive cloud: it contains 5$\times $10$^{4}$ $M_{\sun}$ dense gas as traced by C$^{18}$O \citep{Y}. The cloud also appears to be a rare example of a GMC at an early evolutionary state in that most of the cloud appears cold and not affected by feedback from previous generations of massive star formation \citep{BVC,NBP,HVC}. This makes Vela\,C an excellent target for studying how magnetic fields affect the formation of molecular clouds and dense cloud substructure.

BLASTPol spent 50 hours mapping Vela\,C during the 2012 flight, covering four of the five cloud subregions identified in \cite{HVC}. \cite{FBP} describes the data reduction pipeline, calibration, and polarization de-biasing corrections.  As the telescope beam was non-Gaussian additional smoothing was required to avoid spurious polarization due to sky rotation.  In this paper we use the BLASTPol 500\,$\mu$m data, which required smoothing only to 2\farcm5 FWHM (0.5\,pc) resolution, rather than the 250 or 350 $\mu$m bands, both of which required
smoothing to 3\farcm0 FWHM resolution.\footnote{As noted in \cite{SOLER} the polarization angles are generally consistent between the three BLASTPol bands.}  The resulting 3.1 deg$^2$\,polarization map contains over 1,000 independent polarization measurements.

\cite{FBP} studied the correlations between the 500\,$\mu$m polarization fraction $p$, column density ($N_H$), and the polarization angle dispersion on 0.5 pc scales $S$.  Here we compare our synthetic observations to their sample of interstellar radiation field (ISRF) heated sightlines within the cloud boundaries defined by \cite{HVC}, which excludes sightlines heated by the compact HII region RCW\,36.  In their analysis \cite{FBP} also exclude any sightlines with $p\,<\,0.1\,\%$, $p\,<\,3\,\sigma_p$, or where $p$~varies significantly using different polarized diffuse ISM background subtraction methods. The final catalogue contains 2235 approximately Nyquist sampled sightlines.

\subsection{Numerical Simulations} \label{section:sims}

\subsubsection{Converging Flow Simulations} \label{section:convflow} 

\begin{figure}
\centering
\includegraphics[width=0.86\columnwidth]{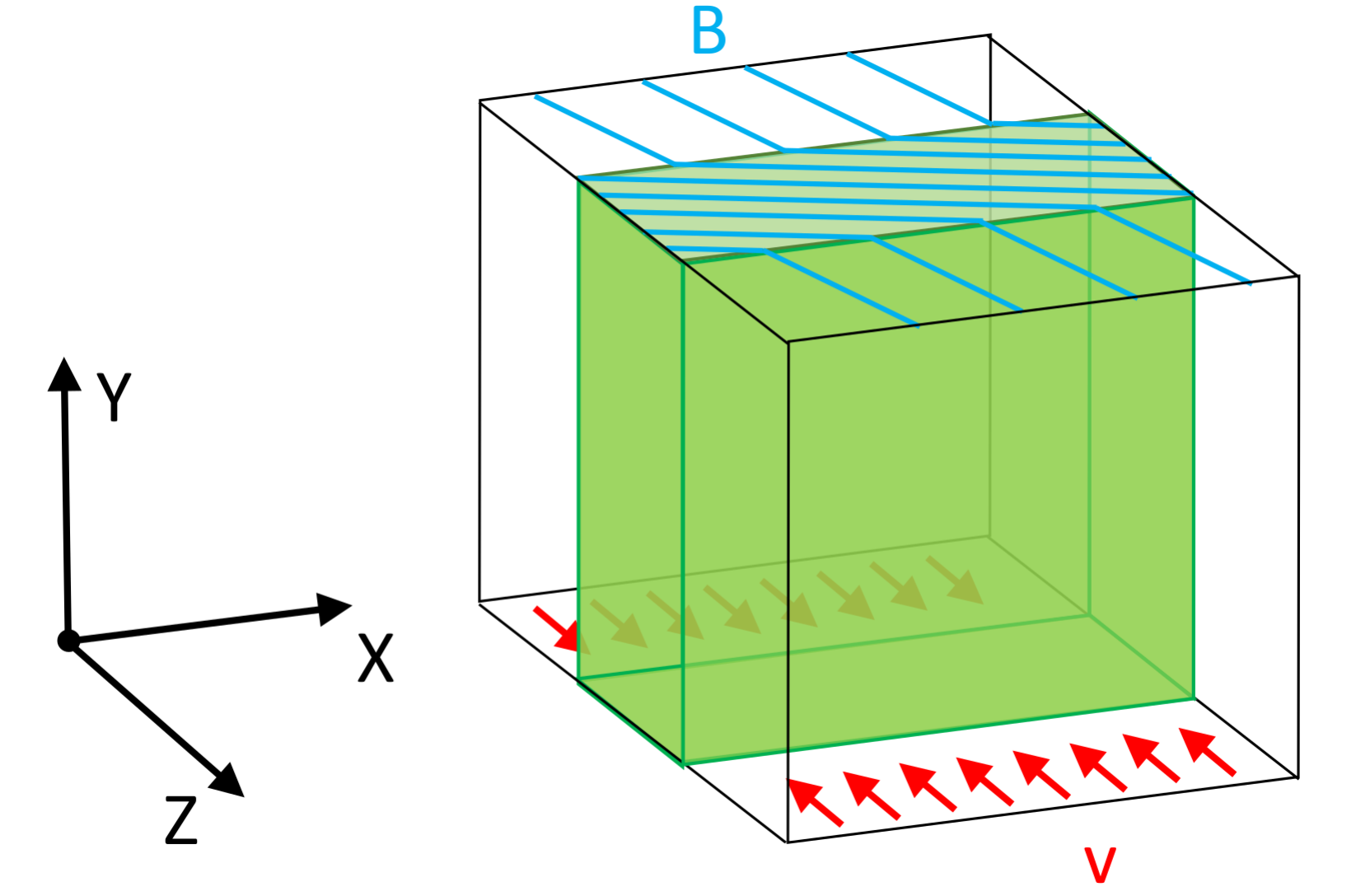}
\caption{Geometry of the \textsc{Athena} converging flow simulations, illustrating the notational conventions labelling the different lines of sight ($x$,$y$,$z$). The converging flows (red arrows) produce a sheet-like post-shock region (green slab); the partially inclined initial magnetic field (cyan lines) becomes amplified during shock compression and results in a prevailing direction in the plane of the post-shock region. The pre-shock region is perturbed by initial turbulence (not indicated for simplicity), which is carried by the converging flow into the post-shock region.}
\label{fig:geom}
\end{figure}

The simulations in this study are similar to those discussed in \citet{CO1,CO2}. These are fully 3-D, ideal MHD colliding flow simulations with gravity conducted using the \textsc{Athena} code \citep{ATHENA}. The colliding flow is adopted as an idealization of the large scale turbulence that is thought to be responsible for driving dense structure formation but is difficult to capture numerically in grid-based simulations like ours that seek to study cloud structures down to the core scale or smaller because of limited dynamic range. An isothermal equation of state is adopted with a sound speed of 0.20 km/s, consistent with the nearly isothermal conditions found in MCs. Initially the simulated region is a uniform density box, with a constant magnetic field in the $x-z$ plane, inclined 20$^{\circ}$ with respect to the $z$ axis. A supersonic, plane-parallel converging flow with a Mach number $\mathcal{M}_s = 10$ is driven in the $\pm z$ direction, which strongly compresses the gas to create a dense post-shock region with a magnetic field amplified by compression. The modestly oblique pre-shock magnetic field becomes flattened and is nearly parallel to the $x-y$ plane in the post-shock region. Material is continuously fed into the simulation through inflow boundary conditions; periodic boundary conditions are adopted on the $x$ and $y$ axes. Turbulence is introduced by perturbing the velocity field, adopting a Gaussian random distribution with a Fourier power spectrum $v_k^2 \propto k^{-4}$ \citep{GO1}. The resulting geometric set-up is illustrated in Figure \ref{fig:geom}.

Using this general set-up, we adopted two simulations chosen to study the role of turbulence and magnetization in shaping the polarimetric observables of the post-shock region. Some important parameters for these two simulations can be found in Table \ref{tab:simpars}. The first model (Model A) used a background density of $\rho_0 = $50 cm$^{-3}$, an initial magnetic field strength of 3.47 $\mu$G, for a simulation box of side-length 10 pc. This model was designed to mimic a large-scale cloud-cloud collision, which has been considered as one of the formation scenarios for denser MCs  \citep{BPHVS,KI,VSRPGG,HHSDB,BVSHK,IF1}. For this model, the velocity perturbation ($\sigma_v = 0.7$ km/s) was chosen by setting the virial number of the simulated cloud, $\alpha_{vir} \equiv (5 R_{cloud}/ G M_{cloud})\sigma_v^2$, where $R_{cloud} \equiv L_{cloud}/2$ and $M_{cloud} \equiv 4\pi R_{cloud}^3 \rho_0/3$ equal to 2. The resulting post-shock region for this model is strongly supersonic ($\mathcal{M}_s = 10.4$) and super-Alfv\'{e}nic ($\mathcal{M}_A = 2.43$). The second model (Model B) adopted was the model M10B10 from \citet{CO2}. The model parameters for this simulation were instead chosen to study a star-forming region inside a magnetized, turbulent dense cloud: a background gas density of 1000 cm$^{-3}$, an initial magnetic field strength of 10 $\mu$G, and a velocity perturbation of 0.14 km/s were chosen for a simulation box of side-length 1 pc for this purpose. The velocity perturbation was chosen by adopting Larson's scaling law for turbulence in MCs, $\sigma_l \propto l^{1/2}$ (see \citet{GO1} or \citet{CO1,CO2} for a detailed derivation). The resulting post-shock region is modestly supersonic ($\mathcal{M}_s = 2.85$) and sub- to trans-Alfv\'{e}nic ($\mathcal{M}_A = 0.81$). To better compare with the BLASTPol observations that are on multi-parsec scales, we rescale this model such that the box side-length is 10 pc (see Section \ref{section:scaling} below.) 

The converging flow geometry and slab-like post-shock regions of these two models provide a unique way to study how intrinsic gas and magnetic structure affects polarimetric observations. An observer looking down each coordinate axis perceives three different conditions with respect to the combined gas-magnetic field structure. An observer whose line-of-sight is in the $z$-direction has a face-on view of the post-shock region with relatively short gas column lengths. In this line-of-sight the mean magnetic field is primarily in the plane-of-sky; only a small component of the magnetic field is parallel to the line-of-sight. An observer whose line-of-sight is in the $y$-direction instead has an edge-on view of the post-shock region with longer gas column lengths, while nevertheless retaining a plane-of-sky oriented mean magnetic field without a strong line-of-sight component. Finally an observer whose line-of-sight is in the $x$-direction has again an edge-on view of the post-shock region but a totally different mean magnetic field orientation, being weak in the plane-of-sky and principally oriented along the line-of-sight. Comparisons between the edge-on and face-on views determine how the amount of material in the column affects polarimetric observations; comparisons between the $x$ line-of-sight and the other two determine how magnetic organization with respect to the observer affects the same. Lastly, comparisons between our two models determine how turbulence and magnetization (or \lq\lq intrinsic\rq\rq~ magnetic disorder) can affect the observations. 

\begin{table*}
	\centering
	\caption{Some properties of the two \textsc{Athena} simulations, including: the initial number density, $n_0$; the pre-shock inflow Mach number $\mathcal{M}_{s,0}$; the turbulent velocity perturbation, $\sigma_v$; the initial magnetic field strength, $B_0$; the initial Alfv\'{e}n Mach number, $\mathcal{M}_{A,0}$; and the post-shock sonic and Alfv\'{e}n Mach numbers $\mathcal{M}_{s,ps}$ and $\mathcal{M}_{A,ps}$; and the post shock plasma $\beta$. The initial densities and magnetic field strength for Model B have been scaled such that the box-length is 10 pc.}
	\label{tab:simpars}
	\begin{tabular}{cccccccccc} %
		\hline
		 Simulation & $n_0$ & $\mathcal{M}_{s,0}$ & $\sigma_v$ & $B_0$ & $\mathcal{M}_{A,0}$ & $\mathcal{M}_{s,ps}$ & $\mathcal{M}_{A,ps}$ & $\beta_{ps}$  \\
		\hline
		Model A & 50.0 cm$^{-3}$ & 10.0 & 0.70 km/s & 3.47 $\mu$G & 2.83 & 10.4 & 2.43 & 0.11 \\
        Model B & 10.0 cm$^{-3}$ & 10.0 & 0.14 km/s & 1.00 $\mu$G & 4.40 & 2.85 & 0.81 & 0.16 \\
		\hline
	\end{tabular}
\end{table*}

\subsubsection{Scaling} \label{section:scaling}

Isothermal, ideal MHD simulations with gravity produce scalable solutions: specific numerical values can be changed under appropriately chosen scaling transformations which leave certain constants and dimensionless numbers unchanged. The inclusion of gravity demands that the numerical value of the gravitational constant $G$ is unchanged under this transformation; this constant has dimensions

\begin{equation}
	[G] = \frac{[v]^2}{[L]^2[\rho]}
    \label{Gdim}
\end{equation}

\noindent where $[L]$ is the dimension of length, $[\rho]$ are the dimensions of density, and $[v]$ are the dimensions of velocity. Additionally, the plasma $\beta = 8\pi \rho c_s^2/B^2$ (the ratio of thermal to magnetic pressure) must also remain unchanged to ensure the magnetic field is scaled appropriately. Plasma $\beta$ has dimensions

\begin{equation}
	[\beta] = \frac{[\rho][v]^2}{[B]^2},
	\label{betadim}
\end{equation}

\noindent where $[B]$ are the dimensions of the magnetic field. In this study it is desirable to leave the velocity unchanged since it is tied to the sound speed, which is controlled by the molecular gas temperature typically fixed at $\sim 10$~K. These constraints and choices provide a complete (though not necessarily unique) scaling transformation for length, density, and the magnetic field using a single multiplicative factor $\lambda$: 

\begin{equation}
L \rightarrow \frac{L}{\lambda},
\end{equation}
\begin{equation}
\rho \rightarrow \lambda^2 \rho, 
\end{equation}
\begin{equation}
B \rightarrow \lambda B.
\end{equation}

As mentioned above, we are free to scale Model B (originally at a box-length of 1 pc) to the same box-length as Model A (whose box-length is 10 pc), provided that the density and magnetic field are adjusted appropriately. These scaling transformations also provide a degree of freedom with respect to the synthetic observations. Column density quantities have dimensions $[N] = [L][\rho]$, indicating that under this transformation they are scaled by $N \rightarrow \lambda N$. Thus we are free to scale the column densities determined from our synthetic observations to values comparable to those determined observationally, provided that we adjust the box-length, number densities, and magnetic field strengths of our simulation. 

\subsection{Synthetic Observations} \label{section:synthpol}

\subsubsection{The Stokes Parameters} \label{section:stokes}

Previous work (e.g., \citealt{LD1,FP1,K1,PXX,CKL}) has established standard practice in the computation of synthetic Stokes parameters from MHD simulations. Suppose we define a Cartesian coordinate system where the $x$ and $y$ coordinates define the plane of the sky (with $y$ corresponding, locally, to Galactic North for definiteness), and with our line of sight $s$ lying parallel to the $z$-axis. Then we may express the synthetic Stokes parameters in terms of the local magnetic field $\mathbf{B} = (B_x,B_y,B_z)$, the source function $S_{\nu}$, and the optical depth $\tau_{\nu}$ \citep{PXX}:

\begin{equation}
I ~= ~~~\int S_{\nu}e^{-\tau_{\nu}} \left(1 - p_0\left(\frac{B_x^2 + B_y^2}{B^2} - \frac{2}{3}\right) \right) ds,
\label{IRT}
\end{equation}
\begin{equation}
Q = p_0 \int S_{\nu}e^{-\tau_{\nu}} \left(\frac{B_y^2 - B_x^2}{B^2}\right) ds,
\label{QRT}
\end{equation}
\begin{equation}
U = p_0 \int S_{\nu}e^{-\tau_{\nu}}\left(\frac{2B_xB_y}{B^2}\right) ds.
\label{URT}
\end{equation}

\noindent ($V = 0$ as thermal dust emission is linearly polarized.) Here, $p_0$ is a parameter called the intrinsic polarization fraction, which is assumed to be uniform over the whole MC. We are comparing to BLASTPol observations at submillimeter wavelengths, so we may safely assume that the emission is optically thin ($\tau_{\nu} << 1$). The source function is usually assumed to be proportional to that of a black-body; as our simulations have adopted an isothermal equation of state we may quote the Stokes parameters in column density units (as opposed to specific intensity units), where $n$ is the local gas number density:

\begin{equation}
I ~ = ~~~ \int n \left(1 - p_0\left(\frac{B_x^2 + B_y^2}{B^2} - \frac{2}{3}\right) \right) ds = N - p_0 N_2,
\label{I}
\end{equation}
\begin{equation}
Q = p_0 \int n \left(\frac{B_y^2 - B_x^2}{B^2}\right) ds ~~~~~~~~~~~~~~~~= p_0\bar{Q},
\label{Q}
\end{equation}
\begin{equation}
U = p_0 \int n \left(\frac{2B_xB_y}{B^2}\right) ds ~~~~~~~~~~~~~~~~~~= p_0 \bar{U}.
\label{U}
\end{equation}

\noindent Here, $N = \int n ~ds$ is the usual definition for column density. Here we have defined the quantities $N_2$, $\bar{Q}$, and $\bar{U}$ as a convenient shorthand:

\begin{equation}
N_2 = \int n \left(\frac{B_x^2 + B_y^2}{B^2} - \frac{2}{3} \right) ds,
\label{N2}
\end{equation}
\begin{equation}
\bar{Q} = \int n \left(\frac{B_y^2 - B_x^2}{B^2} \right) ds.
\label{Qbar}
\end{equation}
\begin{equation}
\bar{U} = \int n \left(\frac{2B_xB_y}{B^2} \right) ds.
\label{Ubar}
\end{equation}

\noindent The $N_2$ term in \eqref{I} is a necessary corrective factor that accounts for the reduction in emission for dust grains inclined with respect to the plane of the sky \citep{FP1}. By examining the extreme configurations that maximize and minimize the contributions in \eqref{N2}, it is clear that this correction safely preserves the condition that Stokes $I$ be strictly positive. $N_2$ ranges from $-\frac{2}{3}N$ (all grains aligned with the line-of-sight) to $\frac{1}{3}N$ (all grains aligned in the plane-of-sky). These extreme configurations should be very rare, and since the correction is also of order $p_0$ - generally a small quantity - then we may assume that Stokes $I$ (in column density units) is approximately the column density to a reasonably high degree of accuracy, an approximation that we confirmed numerically. 

The polarization fraction $p$ is given by

\begin{equation}
p = \frac{\sqrt{Q^2 + U^2}}{I} = p_0 \frac{\sqrt{\bar{Q}^2+\bar{U}^2}}{N - p_0 N_2}.
\label{p}
\end{equation}

\noindent and the polarization angle (measured in the plane of the sky) is given by

\begin{equation}
\chi = \frac{1}{2}\arctan(U,Q),
\label{ch}
\end{equation}

\noindent where $\arctan$ is the two-argument arctangent which returns the appropriate quadrant of the computed angle. Note that $\chi$ is mapped into $[0,\pi)$, as polarization is a pseudovector defining an orientation rather than a direction.

Real observations of the Stokes parameters are limited by the resolution of the instrument. This effect can also be modelled by convolving the pixel-resolution synthetic Stokes parameters with a Gaussian filter. These beam-convolved quantities are then used to compute the polarization fraction and the polarization angle. To examine beam effects we report both the results at pixel-scale and those at telescopic resolution (0.5 parsec) for both Model A and Model B. (Since we have rescaled Model B to the same box-length as Model A, 10 pc, the beam size is the same in both cases.) We implement this beam convolution using a simple symmetric Gaussian filter (implemented by \textsc{scipy} in \citet{SP}). In principle this procedure may be modified to accommodate any beam shape.

Synthetic observations along edge-on lines-of-sight (the $x$ and $y$ lines-of-sight) will have pixels which contain only pre-shock material. In both of our models the post-shock region is the primary region of interest, and the pre-shock material is too diffuse and ordered to properly model an MC. In these lines-of-sight there will also be pixels near the boundary between the pre- and post-shock region; these transition regions will contain sharply varying magnetic fields and artificially short column lengths, potentially contaminating our sample. We exclude these regions simply by eliminating those pixels from the sample, focusing on the interior of the post-shock region.

The intrinsic polarization fraction, $p_0$, is in principle set by conditions in specific MCs. In \cite{FP1}, $p_0$ is explicitly the average (over all grain populations) of the product of two terms: the Rayleigh reduction factor due to imperfect grain alignment, and the reduction in polarization due to the turbulent component of the magnetic field. These conditions are assumed not to vary throughout the cloud\footnote{This assumption is sometimes called \textit{perfect grain alignment}, which is strictly speaking incorrect, as $p_0$ includes the effects of imperfect grain alignment provided that it is the same everywhere. We prefer the term \textit{homogeneous grain alignment.}} and so can be taken safely out of the integrals. We adopt a fiducial value of 0.15 for initial calculations, which is consistent with observationally determined values \citep{PXX,FBP}; however, it bears noting that there is some freedom to adjust this parameter to match observations. Adjusting $p_0$ would not change the polarization angle as the contribution from $p_0$ vanishes in Equation \eqref{ch}.\footnote{In principle, modifications to the grain alignment efficiency that are not homogeneous will introduce some variation in $\chi$, though these corrections might be expected to be small: the corrections will modify both Stokes $Q$ and $U$ and thus will vanish to first order upon computing $\chi$.} The dependence on $p_0$ in Equation \eqref{I} would adjust Stokes $I$ values, but this effect should be small, which we have also verified numerically; in accordance with our approximation that Stokes $I$ is approximately the column density we neglect this effect.

While we are not considering heterogeneous alignment in this paper, a short discussion on its expected effects is useful. In the context of grain alignment by radiative torques \citep{HL1}, grains are expected to be aligned less efficiently with respect to the local magnetic field in denser well-shielded regions. The reduction in alignment efficiency in such regions is expected to decrease the polarization fraction, especially along high column density sightlines. Its effect on the polarization angle along a given sightline is less clear, because the angle is determined by the ratio of the integrated Q and U along the line of sight which, unlike Q and U themselves, does not depend on the grain alignment efficiency in a simple way.  We will explore these effects fully in a subsequent paper.

The synthetic polarimetric observations of the \textsc{Athena} simulations described here were implemented using routines written using \textsc{numpy} \citep{NP}, \textsc{scipy} \citep{SP}, and the \textsc{yt} package \citep{YT}. Our plots were generated using the \textsc{matplotlib} python plotting library \citep{MPL}. \newpage

\subsubsection{Dispersion in Polarization Angles} \label{section:disp}

While the polarization angle $\chi$ provides approximately the column-averaged magnetic field orientation in the plane of the sky, the relative change in orientation rather than its precise value is more directly comparable to observations. This is probed by calculating the dispersion in polarization angles (e.g., \citealt{FG1,PXIX,FBP,CKL}):

\begin{equation}
S^2(\mathbf{x},\delta) = \frac{1}{N}\sum_{i=1}^N \Delta \chi^2(\mathbf{x},\mathbf{x}_i).
\label{S}
\end{equation}

Here, $\Delta \chi(\mathbf{x},\mathbf{x}_i)$ is the angular difference\footnote{Note that because polarization angles $\pi$ out of phase are indistinguishable, the maximum angular difference must be $\pi/2$.} between the angle at the point $\mathbf{x}$ and another point $\mathbf{x}_i$ located a distance $\delta$ (called the \textit{lag}) away from it. The sum is over all points at the lag radius $\delta$ away from $\mathbf{x}$. $\Delta\chi_i$ is usually calculated directly from the Stokes parameters using the two-argument arctangent \citep{PXIX}:

\begin{equation}
\Delta \chi(\mathbf{x},\mathbf{x}_i) = \frac{1}{2}\arctan(Q_iU_{\mathbf{x}} - Q_{\mathbf{x}}U_i,Q_iQ_{\mathbf{x}}+U_iU_{\mathbf{x}}),
\end{equation}

\noindent where $Q_{\mathbf{x}}$ and $U_{\mathbf{x}}$ are the Stokes parameters at $\mathbf{x}$ and $Q_i$ and $U_i$ are the Stokes parameters located at the point $\mathbf{x}_i$ located a distance $\delta$ away from $\mathbf{x}$.

The dispersion in polarization angles is a measure of the local changes in magnetic field direction, regardless of orientation convention chosen. We note that, as calculated in \citet{PXIX}, the dispersion in polarization angles for pure noise converges to $\pi/\sqrt{12}$ (about 52$^{\circ}$). For our purposes we elect to use the smallest sensible lag to study the finest polarization angle structure. For BLASTPol and the beam-convolved synthetic observations, the lag would be the FWHM of the beam; we use the pixel scale for the simulation resolution synthetic observations. 

\begin{figure*}
\includegraphics[width=0.90\textwidth]{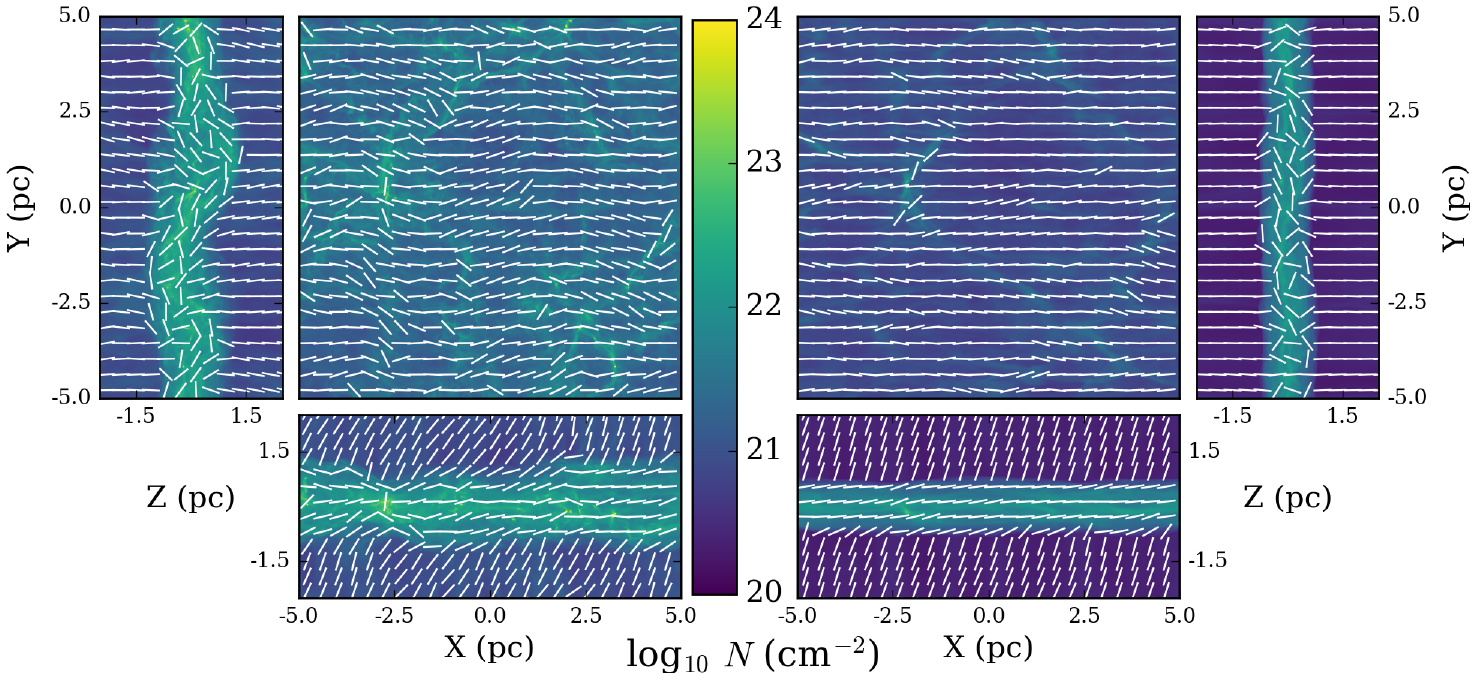}\\
\includegraphics[width=0.90\textwidth]{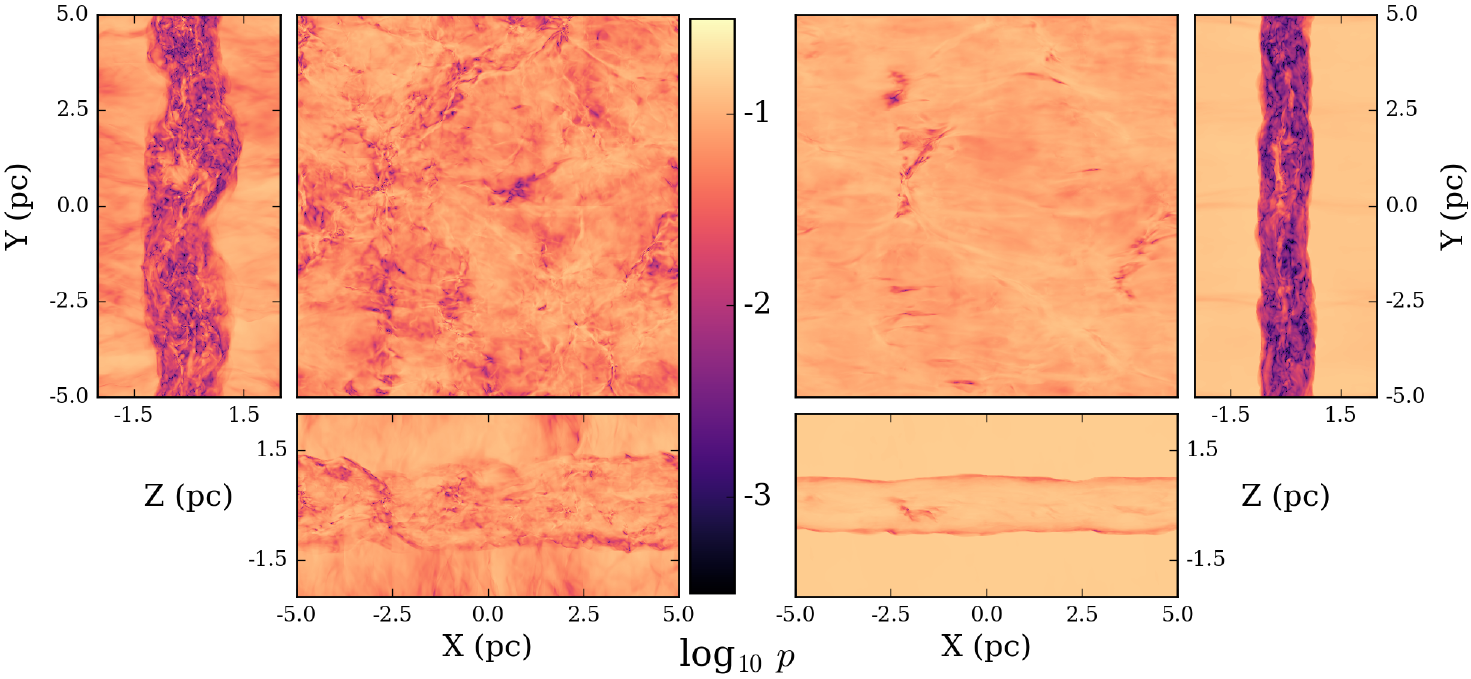}\\
\includegraphics[width=0.90\textwidth]{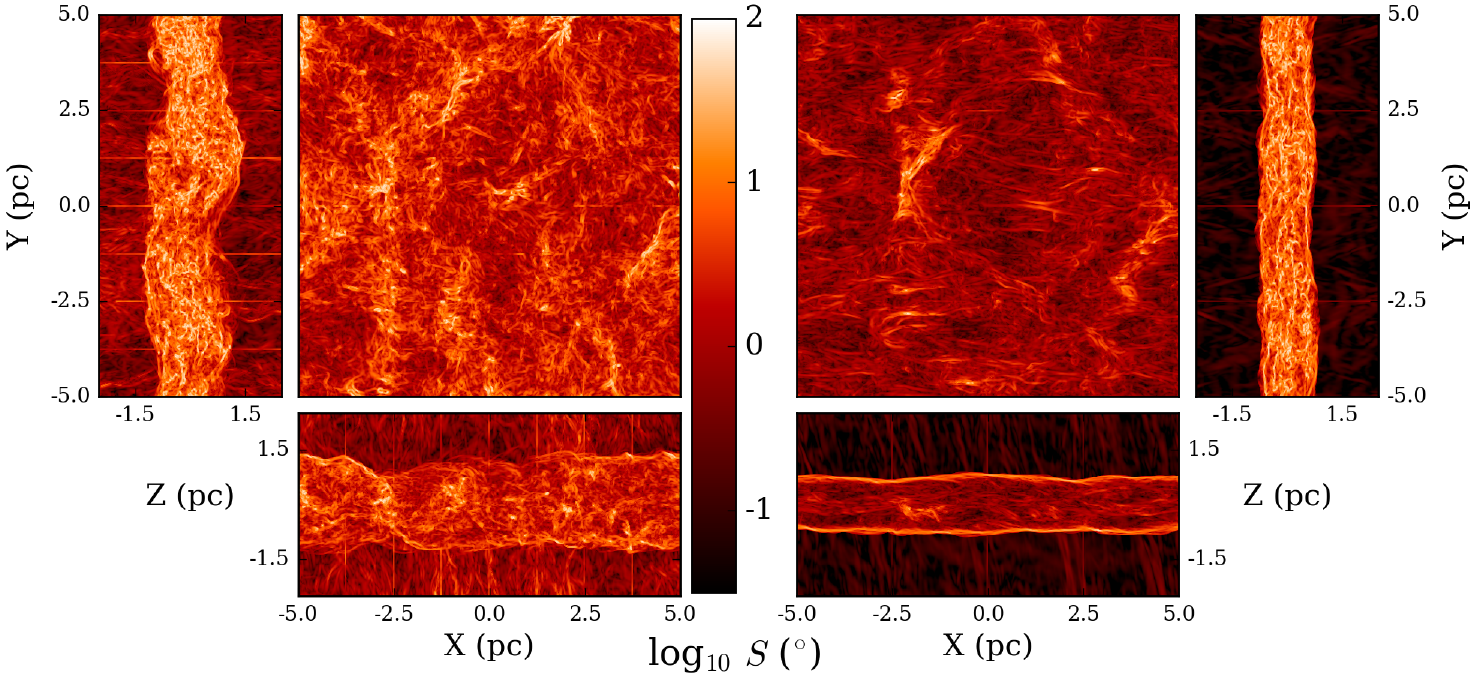}
\caption{Stokes $I$ (top row), polarization fraction (middle row), and dispersion in polarization angles (bottom row) for the Athena Model A simulation (left column) and the Athena Model B simulation (right column). For each image, the central image is the $z$ line-of-sight; the image parallel to the $y$-axis is the $x$ line-of-sight; and the bottom image parallel to the $x$-axis is the $y$ line-of-sight. These images are produced at pixel resolution.}
\label{fig:simpics}
\end{figure*}

\begin{figure*}
\includegraphics[width=0.90\textwidth]{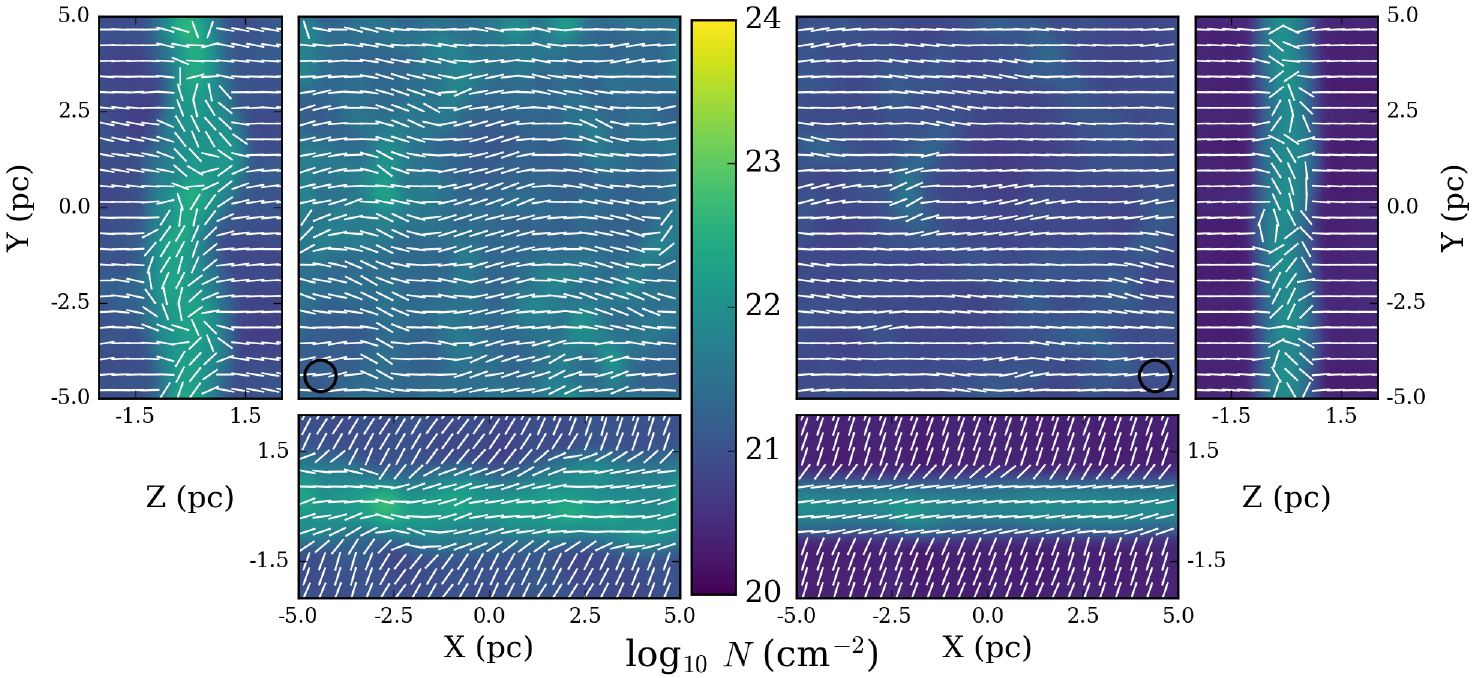}\\
\includegraphics[width=0.90\textwidth]{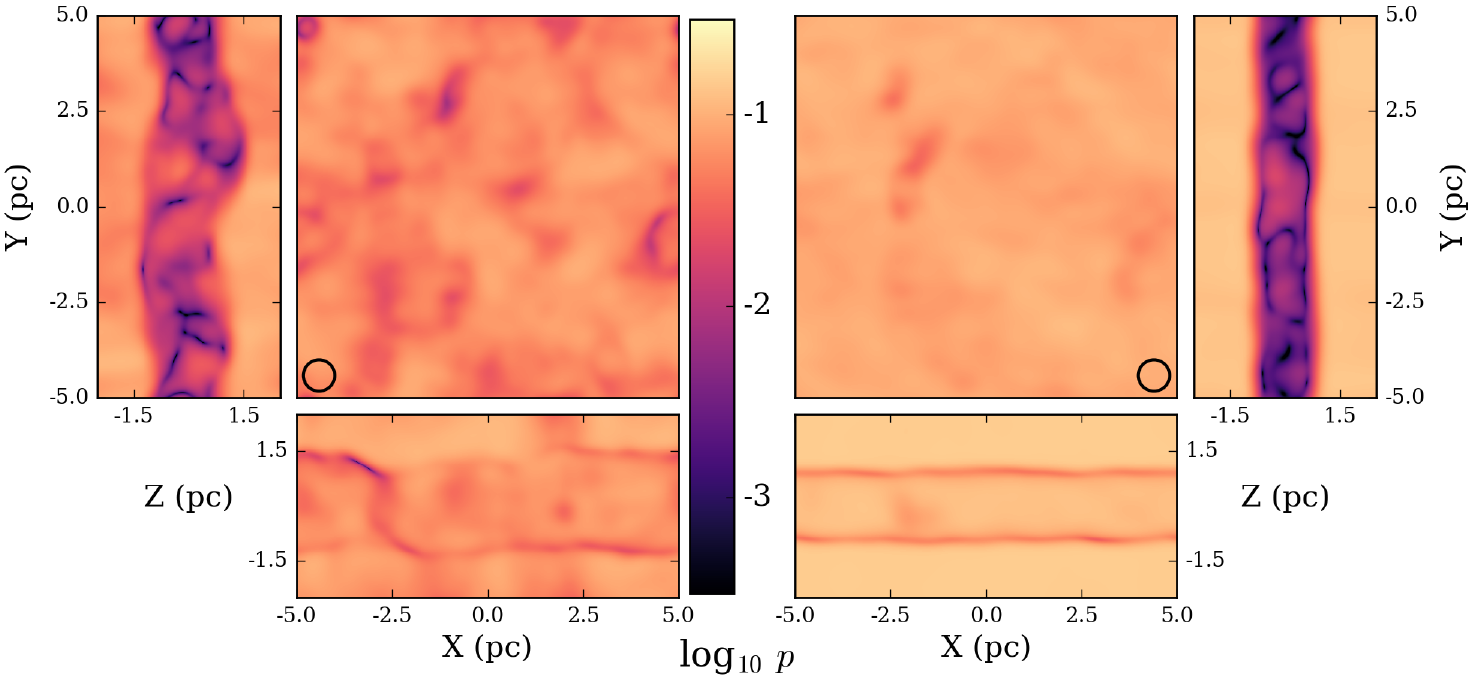}\\
\includegraphics[width=0.90\textwidth]{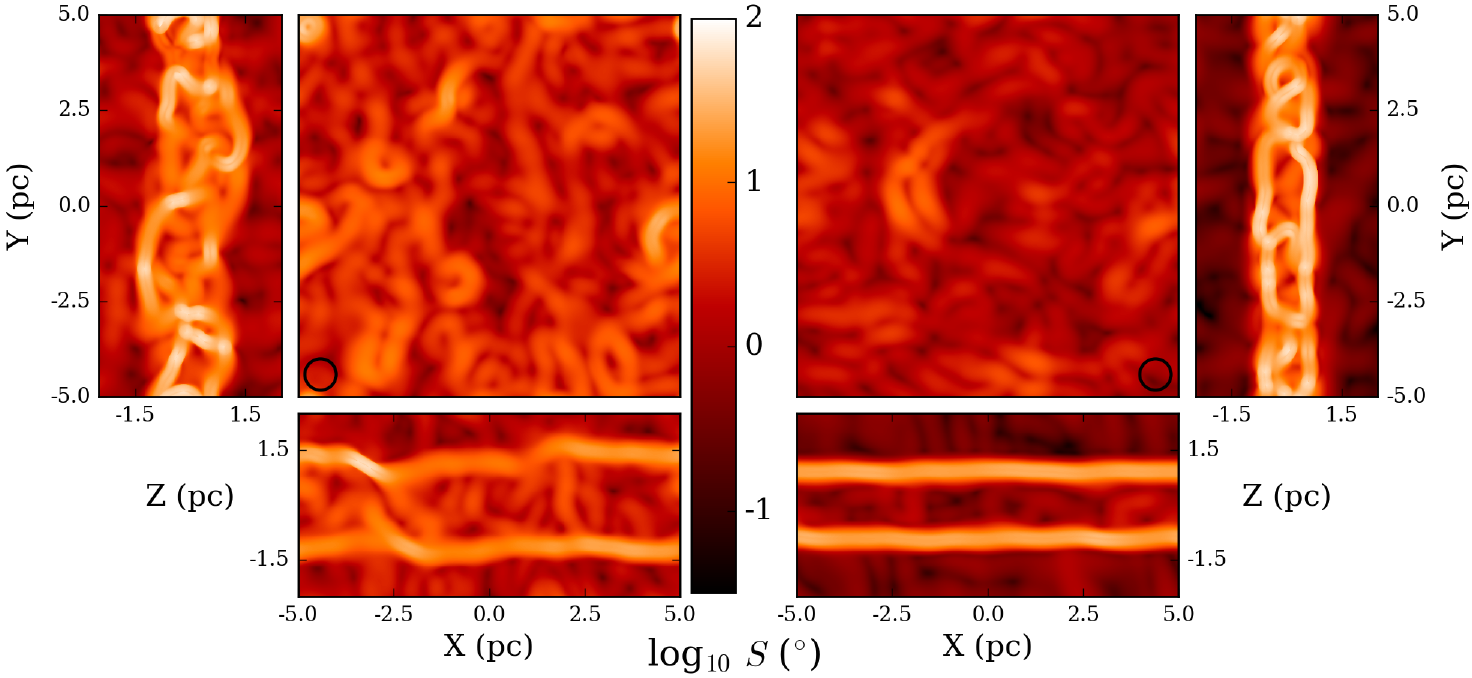}
\caption{Same as Figure \ref{fig:simpics} but convolved with the annotated 0.5 pc beam. $p$ and $S$ are computed from the beam convolved Stokes parameters ($Q$ and $U$; see Section \ref{section:synthpol}). A black circle annotates the beam FWHM in each plot.}
\label{fig:simpicsBM}
\end{figure*}

Maps of the column density, polarization fraction, and dispersion in polarization angles computed along the $x$, $y$, and $z$ lines-of-sight are presented in Figure \ref{fig:simpics} (at pixel resolution) and in Figure \ref{fig:simpicsBM} (convolved with a Gaussian beam.) The polarization angles (corresponding to the magnetic field orientation) are annotated on the column density plots (top panel). 

\subsection{Statistical Techniques} \label{section:stats}

\subsubsection{Geometric Statistics} \label{section:logstats}

The range of values for column density, polarization fraction, and the dispersion in polarization angles typically spans several orders of magnitude \citep{VSG,FBP,PXIX}; studying their probabilistic features is commonly done in logarithmic space to capture both central and asymptotic behaviour in the distribution. For the column density and polarization fraction, we will work in logarithmic contrast variables, normalizing the values to a measure of central tendency:

\begin{equation}
\zeta_{N_H} = \log_{10}\left(\frac{N_H}{\overline{N_H}}\right),
\end{equation}
\begin{equation}
\zeta_{p} ~~~= \log_{10}\left(\frac{p}{\overline{p}}\right).
\end{equation}

\noindent For the dispersion in polarization angles, we will instead simply use the value measured in degrees: 

\begin{equation}
\zeta_{S} ~~~= \log_{10}\left(\frac{S}{1^{\circ}}\right).
\end{equation}

In logarithmic space, the geometric mean is a more natural measure of central tendency than the arithmetic mean. For a set of $N$ values $X = \{x_i\}$,  the geometric mean is simply related to the arithmetic mean of the logarithmic values, defined by $Z = \{\zeta_i\} = \{\log x_i\}$:

\begin{equation}
\mu_G(X) = \left(\prod_{i=1}^N x_i\right)^{1/N} = \exp\left(\frac{1}{N} \sum_{i=1}^N \zeta_i \right) = \exp\left(\mu_A(Z)\right),
\end{equation}

\noindent where $\mu_G$ and $\mu_A$ denote the geometric and arithmetic means of the set, respectively. Similarly, the geometric standard deviation is a natural first moment measure of distribution width in logarithmic space, and is defined through the standard deviation of the logarithmic contrast values \citep{GEOSTAT}:

\begin{equation}
(\log \sigma_G(X))^2 = \frac{1}{N}\sum_{i=1}^N \log\left(\frac{x_i}{\mu_G(X)}\right)^2 = \sigma^2(Z),
\end{equation}

\noindent where $\sigma_G$ and $\sigma$ denote the geometric standard deviation and standard deviation of a set, respectively. (These expressions are easily modified to base 10.)

Higher statistical moments can provide more information on the shape of a probability distribution. The kurtosis of a distribution (the fourth standardized moment), in particular, can offer insight into the behaviour of the tails of the distributions \citep{SDMMLA}. To study the tails of the distribution in logarithmic space, we may compute the kurtosis of the logarithmic values of the polarimetric observables, which we call the \textit{geometric kurtosis}:\footnote{Rather than adopting the same convention in the geometric mean and geometric standard deviation, we define the geometric kurtosis not as $\exp(\text{Kurt}(Z))$ but as $\text{Kurt}(Z)$, as the prior definitions are chosen to emphasize the connection between geometric moments of the set and the arithmetic moments of the logarithm of the set. No simple relationship exists for the higher statistical geometric and arithmetic moments.}

\begin{equation}
\text{Kurt}_G(X) = \text{Kurt}(Z) = \frac{\frac{1}{N}\sum_{i=1}^N\left(\log\left(\frac{x_i}{\mu_G(X)}\right)\right)^4}{\log(\sigma_G(X))^4}
\end{equation}

\noindent As a Gaussian distribution has a kurtosis of 3, the excess kurtosis (the kurtosis less 3) is often quoted to emphasize deviations from Gaussianity \citep{SDMMLA}. (We denote excess kurtosis by $\overline{\text{Kurt}}$ to distinguish it.) Similarly, we will quote the excess geometric kurtosis, which describes deviations from log-normality. Distributions with positive excess kurtosis are termed \textit{leptokurtic} and have tails which asymptotically approach zero less rapidly than a Gaussian; distributions with negative excess kurtosis are instead termed \textit{platykurtic} and have tails that approach zero more rapidly than a Gaussian distribution. Distributions with excess kurtosis close to zero are termed \textit{mesokurtic.} In our logarithmic context, power-law asymptotics (a commonly encountered behaviour) would manifest as positive excess geometric kurtosis. 

Both the freedom to adjust $N$ values through a scaling transformation and the freedom to adjust $p_0$ are multiplicative factors, and therefore in logarithmic space amount to adjusting the geometric mean. These adjustments will not adjust either the geometric standard deviation nor the geometric kurtosis. For these reasons, the geometric mean of the polarization fraction and column density are less important quantities as far as constraining simulation parameters is concerned. However, because the dispersion in polarization angles cannot be scaled freely, the mean angle dispersion is an important quantity of interest.

\subsubsection{Kernel Density Estimation} \label{section:KDE}

The underlying probability density function (PDF) for the observables has much to say about molecular cloud structure. For a single observable, the PDF is commonly visualized using a histogram. A two-dimensional version of the same is often used when studying joint correlations. The histogram has the advantage of simplicity, but has its limitations. Regardless of how optimally the bin size is chosen, information is lost inside the bin. Locating the bin centres and choosing bin size given a dataset is arbitrary, and behaviour at the bin boundaries can give rise to inappropriate discontinuities \citep{KDE,FB}. Instead we prefer to use kernel density estimation (KDE), a technique that yields a smooth estimate of the PDF from a set of data. 

For a set $\{x_i\}$ of $N$ independent observations which are sampled from a common PDF $f$, the kernel density estimate of $f$ is given by \citep{KDE}

\begin{equation}
\hat{f}_h(x) = \frac{1}{N}\sum_{i=1}^N \frac{1}{h(x)} K\left(\frac{x - x_i}{h(x)}\right).
\label{KDE}
\end{equation}

\noindent Here $K$ is the kernel function (normalized and with mean zero) and $h$ is the bandwidth parameter. $h$ may be thought of as analogous to the bin size in histograms. It is chosen according to Scott's rule, a known rule for optimal bandwidth \citep{KDE}. The choice of kernel function is usually unimportant, and may depend on the application. In our case, it has been established \citep{VSG} that a log-normal distribution is expected for the column density. Therefore, we use a Gaussian kernel function on the logarithmic values:

\begin{equation}
K(u) = \frac{1}{\sqrt{2\pi}}e^{-\frac{1}{2}u^2}.
\end{equation}

\noindent These one-dimensional forms are easily generalized to two or higher dimensions for multivariate correlations. For a set $\left\{\mathbf{x}_i\right\}$ of $N$ independent observation vectors of dimension $d$, the kernel density estimate of the joint probability density function $f(\mathbf{x})$ is \citep{KDE} 

\begin{equation}
\hat{f}_{\mathbfss{H}}\left(\mathbf{x}\right) = \frac{1}{N}\sum_{i=1}^N \left|\mathbfss{H}\right|^{-1/2} K(\mathbfss{H}^{-1/2}\left(\mathbf{x} - \mathbf{x}_i\right)),
\end{equation}

\noindent where $\mathbfss{H}$ is the multivariate generalization of the bandwidth parameter (the $d\times d$ bandwidth matrix) which may also be chosen optimally. The multivariate Gaussian kernel function is

\begin{equation}
K(\mathbf{u}) = \frac{1}{(2\pi)^{d/2}}e^{-\frac{1}{2}\mathbf{u}^T \mathbf{u}}.
\end{equation} 

\noindent We use the \textsc{scipy} implementation of Gaussian KDE \citep{SP}, which implements both the univariate and multivariate cases.

\subsubsection{Principal Components and the Covariance Matrix} \label{section:PCA}

The joint PDF of observables contains strictly more information than the one dimensional PDFs alone, which may be thought of as projections of the full configuration space PDF onto a single observable axis. The joint distributions contain not only information on extent but also on mutual dependence. Past work \citep{PXIX,PXX,FBP} has presented these joint correlations and explored mutual dependence by fitting to a power law with linear regression. Linear regression has the disadvantage of depending on the choice of independent variable, i.e. the fit is not symmetric with respect to axis choice. Linear regression is one choice in characterizing the geometry of the multivariate joint distribution; we instead compute the principal components for this purpose \citep{MH}. Principal Component Analysis (PCA) has been used elsewhere in astrophysics, where it has particular value in reducing the dimensionality of datasets with many variables, such as is done in \citet{PCA}. Unlike these previous applications, our use of the principal components is based on their natural geometric interpretation and utility as descriptive tools. 

Suppose that you have a set of $n$ variables ${X_n}$, each of which is a set of observations of length $N$. The covariance between two variables is given by

\begin{equation}
\sigma^2\left(X_i,X_j\right) = \displaystyle\frac{1}{n - 1}\sum_{k=1}^n (X_{ik} - \bar{X_i})(X_{jk} - \bar{X_j}).
\label{COV}
\end{equation}

\noindent Here, $\bar{X_i}$ refers to the average of $X_i$. This definition recovers the ordinary definition of variance:

\begin{equation}
\sigma^2(X_i) = \displaystyle\frac{1}{n-1}\sum_{k=1}^n (X_{ik} - \bar{X_i})^2.
\label{V}
\end{equation}

\noindent The covariance matrix for this set of variables is this symmetric $n\times n$ matrix:

\begin{equation}
\mathbfss{C} = \left( 
			   \begin{array}{ccccc}
               \sigma^2(X_1)     & \cdots & \sigma^2(X_i,X_1) & \cdots & \sigma^2(X_n,X_1) \\
               \vdots            & \ddots &                   &        &                   \\
               \sigma^2(X_1,X_i) &        & \sigma^2(X_i)     &        & \vdots            \\
               \vdots            &        &                   & \ddots &                   \\
               \sigma^2(X_1,X_n) &        & \cdots            &        & \sigma^2(X_n)     \\
               \end{array}
			   \right).
\label{COVM}
\end{equation}

\noindent The principal components are the eigenvectors $\{\mathbf{v}_m\}$ of $\mathbfss{C}$, satisfying:

\begin{equation}
\mathbfss{C}~\mathbf{v}_m = \lambda_m \mathbf{v}_m.
\label{EV}
\end{equation}

\noindent Trivially there are $n$ principal components, mutually orthogonal to each other in the $n$-dimensional configuration space of our variables. Analogous to the principal moments of inertia for a rotating rigid body, the eigenvectors describe a set of coordinate axes that maximize variance (in the least-squares sense.) Principal axes, in this sense, have been used effectively to describe the shape of clumps in MCs \citep{GLSO,NL,GO2}. The eigenvalues describe the relative importance of the principal components: the component with the highest eigenvalue contains the most variance (and is often referred to simply as \textit{the} principal component, especially for multivariate studies with large $n$.) We note that PCA provides a very simplified view of the joint PDF geometry, and is unsuited for detailed studies of PDF features, which would require more sophisticated techniques. Analogous to the fitted slope in linear regression, the implied PCA power law index is the slope of the principal vector. 

The principal components capture the geometry of the joint distributions, but do have an important drawback: they do not measure, without ambiguity, the degree to which observables are correlated with each other. A simple measure of correlation between two observables is the Pearson correlation coefficient \citep{FB}. For two observables, $X$ and $Y$, their Pearson correlation coefficient is \citep{FB}

\begin{equation}
\rho_{P,XY} = \frac{\sigma^2(X,Y)}{\sqrt{\sigma^2(X) \sigma^2(Y)}}.
\label{PCC}
\end{equation}

\noindent It should be noted that (like PCA) the Pearson correlation coefficient is limited in its sensitivity to correlations more complicated than simple linear ones. A slightly more sophisticated version, the Spearman rank correlation coefficient, tests for monotonic dependence rather than simple linear correlation. If $R_X$ and $R_Y$ are the ranked variables\footnote{The ranked variables are the integer ordering of the observations, with the largest value assigned to be 1. Duplicate values are provided a fractional rank.} corresponding to $X$ and $Y$, then the Spearman rank correlation coefficient is the Pearson correlation coefficient of these two ranked variables \citep{FB}:

\begin{equation}
\rho_{S,XY} = \frac{\sigma^2\left(R_X,R_Y\right)}{\sqrt{\sigma^2\left(R_X\right) \sigma^2\left(R_Y\right)}} .
\label{SCC}
\end{equation}

\noindent The numerical value of $\rho_{P,XY}$ and $\rho_{S,XY}$ is between -1 and 1, where positive values indicate a positive correlation, and negative values indicate negative correlation. The magnitude is a measure of correlation strength, with 1 being associated with perfect linear correlation and 0 being perfectly uncorrelated. For our purposes we compute the Pearson correlation coefficient of the logarithmic values of the observables (equivalent to applying a logarithmic transformation on the data.) Therefore this Pearson coefficient measures how strongly the two observables are relatable to each other via a power law. As a logarithmic transformation is monotonic, the Spearman rank correlation coefficient is unchanged by a logarithmic transformation; therefore this quantity measures the degree of monotonic dependence, without imposing a power-law form onto the relationship.

\section{The Polarization Fraction} \label{section:p} 

\begin{figure*}
	\includegraphics[width=2.1\columnwidth]{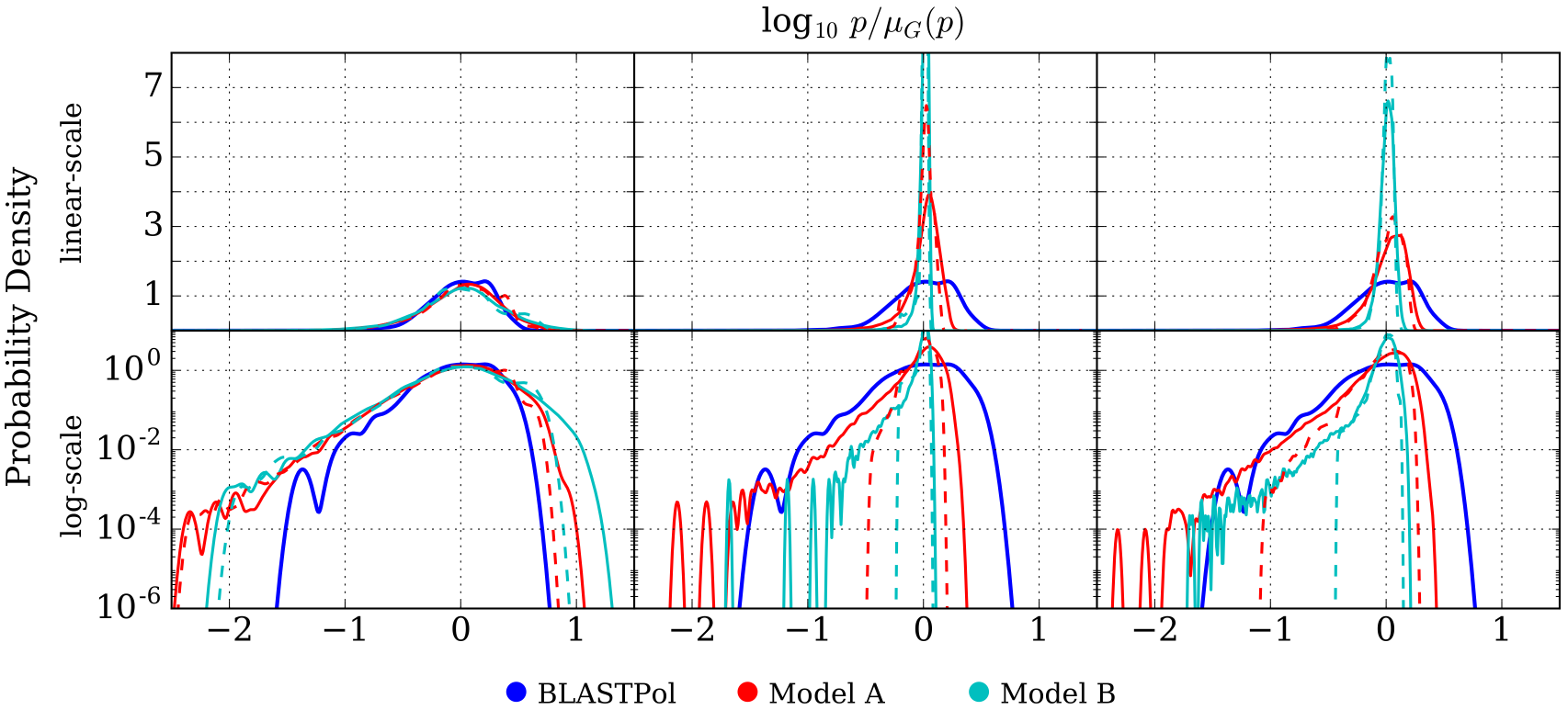}
    \caption{PDFs of the polarization fraction for the BLASTPol Vela C observations (\citealt{FBP}, Blue), Model A (Red), and Model B (Cyan). The solid coloured lines correspond to distributions computed at the pixel scale of the simulation, while dashed coloured lines correspond to those observed with a Gaussian beam. The $x$ line-of-sight distributions are in the left column; in the centre, the $y$ line-of-sight distributions; and in the right, the $z$ line-of-sight distributions. The top row PDFs are log-linear and the bottom row are log-log.}
    \label{fig:pPDF}
\end{figure*} 

\begin{table}
	\centering
	\caption{Distribution statistics for the one-dimensional PDFs of the polarization fraction.}
	\label{tab:pStats}
	\begin{tabular}{cccccc} %
		\hline
		 LOS & Model & Beam & $\mu_G(p)$ & $\log\sigma_G(p)$ & $\overline{\text{Kurt}}_G(p)$  \\
		\hline
		$x$-LOS & A & Pixel   & 0.017 & 0.330 & 1.214 \\
                &   & 0.5 pc  & 0.009 & 0.323 & 1.409 \\
		        & B & Pixel   & 0.007 & 0.368 & 0.924 \\
                &   & 0.5 pc  & 0.003 & 0.350 & 0.903 \\
        $y$-LOS & A & Pixel   & 0.068 & 0.153 & 8.987 \\
                &   & 0.5 pc  & 0.062 & 0.088 & 1.252 \\
		        & B & Pixel   & 0.125 & 0.066 & 49.34 \\
                &   & 0.5 pc  & 0.122 & 0.045 & 6.469 \\
        $z$-LOS & A & Pixel   & 0.061 & 0.195 & 5.997 \\
                &   & 0.5 pc  & 0.057 & 0.152 & 3.814 \\
		        & B & Pixel   & 0.091 & 0.085 & 29.85 \\  
                &   & 0.5 pc  & 0.090 & 0.061 & 5.805 \\
        \textbf{BLASTPol} & - & 0.5 pc & \textbf{0.027} & \textbf{0.260} & \textbf{0.737} \\
		\hline
	\end{tabular}
\end{table}

The polarization fraction measured at any pixel is strongly dependent on the magnetic organization within its line-of-sight. On the one hand, contributions to the polarized emission at any point in the line-of-sight are bounded above by the inclination of the magnetic field with respect to the plane-of-sky: any inclination reduces the polarization signal by reducing the apparent ratio of the long to short axes of the grain relative to the observer \citep{CKL}. On the other hand, even contributions with no inclination can be negated entirely by another contribution exactly $\pi/2$ out of phase, or partially negated by any contribution with non-zero relative phase. This makes the polarization fraction a simultaneous measure of both magnetic field inclination with respect to the plane-of-sky and magnetic field organization along the line-of-sight (provided that contributions from heterogeneous grain alignment can be neglected.) Examining the distribution of polarization fraction thus provides some insight into the general behaviour of the magnetic field as seen by the observer. 

While we noted in Section \ref{section:synthpol} that there is freedom in the choice of $p_0$, there is not unlimited freedom: $p_0$ is limited to attain values consistent with both grain alignment physics and the amount of magnetic disorder at scales smaller than the simulation resolution element. Practically $p_0$ is determined empirically from the polarization fraction distribution (effectively, from the mean polarization fraction.) In principle, an overall reduction in the mean polarization fraction from sky-averaged inclination and cancellation along all lines-of-sight would also reduce the apparent $p_0$ from magnetohydrodynamical effects alone. We find, however, that these effects are tied to the mean orientation of the magnetic field with respect to the line-of-sight. The mean polarization fractions are found in Table \ref{tab:pStats}. (These means are all computed after assuming the fiducial value $p_0 = 0.15$.) In the $y$ and $z$ lines-of-sight, we find that the mean polarization fraction for Model A is lower than that for Model B. In this case, the disordering influence of stronger turbulence/weaker magnetization reduces the mean polarization fraction. On the other hand, in the $x$ line-of-sight, the mean polarization fraction for Model A is higher, not lower, than Model B. In this case, the more disordered magnetic field relative to the mean magnetic field provides more opportunities along any given line-of-sight for the magnetic field to locally align with the plane-of-sky, strengthening contributions to the polarization fraction. The lower turbulence and stronger magnetization of Model B permits fewer deviations from the mean magnetic field, resulting in a more complete suppression of polarized emission when the mean magnetic field is parallel to the line-of-sight. 

Turning our attention to the PDFs of polarization fraction contrast in Figure \ref{fig:pPDF}, we can see immediately that the $x$ line-of-sight shows remarkable agreement with the BLASTPol observations of Vela C. Distribution widths and kurtosis can be found in Table \ref{tab:pStats}. The $x$ line-of-sight distribution widths for the simulations and BLASTPol are all relatively similar (left panel of Figure \ref{fig:pPDF}), and agreement improves slightly upon beam convolution (dashed lines in Figure \ref{fig:pPDF}). In terms of kurtosis, both BLASTPol and the simulations are generally mesokurtic (deviating little from Gaussian). The leading edges at high polarization fraction contrast are also generally consistent, as are the tails at low polarization fraction. In contrast, neither the $y$ or $z$ lines-of-sight (middle and right panels of Figure \ref{fig:pPDF}) produce PDFs that are very consistent with the BLASTPol observations. These lines-of-sight produce significantly narrower distributions than the BLASTPol observations, and are all significantly leptokurtic, with extremely peaked distributions that have more extended tails than a Gaussian. This translates to a very uniform polarization fraction in the plane-of-sky, with significant deviations being quite rare. The leading edges are much steeper, and the tails at low polarization fraction are generally lower in probability than the BLASTPol tail; such steep leading edges indicate that the peak is probably very near the maximum polarization fraction (see, e.g., the log-log plots of the $y$ and $z$ lines-of-sight in Figure \ref{fig:pPDF}, bottom panel). 

Examining the $y$ and $z$ lines-of-sight in more detail, we note their remarkable consistency: viewing either Model A or Model B from either line-of-sight produces little variation. Both of these lines-of-sight differ primarily in that one is edge-on ($y$) versus the other being face-on ($z$), which indicates that the $y$ line-of-sight sightlines generically contain much more material than the $z$ line-of-sight. Because both of these lines-of-sight share the quality that the mean magnetic field is primarily in the plane-of-sky and perpendicular to the line-of-sight, this similarity indicates that the magnetic field orientation dominates the behaviour of the polarization fraction, and is far more important than the effects that might arise due to longer column integration lengths. 

This is an important observation: consider the aforementioned roles of inclination and cancellation within a column. Under this regime, where the mean magnetic field is in the plane-of-sky, then a longer column of material would provide more opportunities to encounter fluctuations relative to the mean magnetic field. The effects of more fluctuations are best understood relative to the mean polarization fraction. On the one hand, if more fluctuations can be encountered, then intuitively it is less likely for high polarization fraction contrast to manifest, as such configurations demand significant magnetic order in the column to avoid cancellation. On the other hand, the lowest polarization fractions also become less likely. This perhaps counter-intuitive result arises due to the conditions necessary to achieve such low polarization fractions: in this regime, consistently high inclination (where the magnetic field is nearly aligned with the line-of-sight) throughout the column is very unlikely, and therefore low polarization fractions require significant pairwise cancellation. While this indicates a state of extremely high magnetic disorder, it is also a very rare state. The likelihood of achieving such a state is improved if there is less material required to pairwise cancel; or equivalently, higher rates of fluctuations can destroy the pairwise cancellation, simply by serendipitous magnetic field orientation alignment between two fluctuations. The combined effect is that, magnetic organization being the same, longer column lengths result in narrower polarization fraction distributions. This is corroborated in Figure \ref{fig:pPDF} (middle and right panels), where we see narrower distributions in the $y$ line-of-sight relative to the $z$ line-of-sight. However, it is important to note that the relative weakness of this effect further demonstrates the importance of magnetic organization relative to the observer.

Within these lines-of-sight, the differences between Model A and Model B are apparent. Model A has a wider distribution, has less kurtosis, has reduced steepness in the leading edge, and has a higher probability state for the depolarization tail. These differences suggest that stronger turbulence/lower magnetization (Model A vs. Model B) has a role in determining the shape of the polarization fraction distribution, at least in the presence of an ordered mean magnetic field in the plane-of-sky. Both the increase in width and decrease in kurtosis indicate that turbulence tends to counteract the highly peaked behaviour. This is consistent with the general expectation for the effects of turbulence, which would introduce disorder and weaken signatures of strong magnetic order. 

The $x$ line-of-sight, as well as the BLASTPol observations, both display significantly different behaviour from that of the $y$ and $z$ lines-of-sight. There is not a particularly distinctive peak, and the transition to the depolarization tail is indistinct. Notably, there is remarkably little difference in polarization fraction contrast between either Model A or Model B despite their different levels of turbulence and magnetization. The only meaningful difference appears to be at highest polarization fraction levels, wherein Model B has a slight enhancement at high polarization fraction relative to Model A. This could possibly be due to the increased role of turbulence in Model A: stronger turbulence and lower magnetization could decrease the likelihood of attaining the highest polarization fraction contrast. Because in the $x$ line-of-sight, contributions to the Stokes parameters (and thus the polarization fraction) from the mean magnetic field are suppressed, the highly ordered high polarization fraction configurations along a column become very unlikely. Increasing the turbulence (or making the magnetic field less resistant to perturbation) could reduce this likelihood further, effectively narrowing the distribution in the tails. Much like what we see with the mean polarization fraction, the influence of turbulence/magnetization is significantly tied to the orientation of the mean magnetic field with respect to the observer: in the $y$ and $z$ line-of-sight, we see that stronger turbulence/lower magnetization widens the distribution, yet in the $x$ line-of-sight, this narrows it. 

The beam convolved distributions are provided as dashed lines in Figure \ref{fig:pPDF}. The effects of beam convolution are most concentrated in the depolarization tails and the leading edge, where the highest and lowest polarization fraction contrasts appear to be cut off or sharply curtailed. Note that beam convolution does not broaden distributions, but narrows them: the beam mixes information spatially on the plane-of-sky, and therefore any contrast features smaller than the beam will be partially destroyed. In the case of the $y$ and $z$ lines-of-sight, we see that the depolarization tail nearly disappears and the leading edge steepens further. The peak becomes even more emphasized and narrow, though the kurtosis is reduced (most likely due to the reduction in the depolarization tail.) This demonstrates that the scale of the highest and lowest contrast features in these lines-of-sight are not larger than the beam and probably not clustered enough to avoid destruction after beam convolution, or else they would be preserved. On the other hand, in the $x$ line-of-sight the leading edge is reduced, but not the depolarization tail. Since the tail is preserved under beam convolution, we may infer that either the scale of these features is larger than the beam or that they are ubiquitous enough to avoid destruction. 

The depolarization tails in all the polarization fraction distributions demonstrate interesting asymptotic behaviour. This includes the BLASTPol distribution, though it appears to be cut off at the lowest polarization fractions. This is likely a result of the finite polarization sensitivity of the BLASTPol telescope, which necessarily cannot detect extremely low polarization fractions.\footnote{As noted in Section \ref{section:BP}, \citet{FBP} only includes sightlines with $p > 0.1\%$. Lower polarization values suffer from uncertainty in correcting for the instrumental polarization of the telescope.} In all the distributions, the leading edge rapidly falls off from the peak, as evidenced in the log-log plot in Figure \ref{fig:pPDF} (bottom panel). This is consistent with Gaussian behaviour, in which probability is exponentially attenuated at high contrast values. Instead, the depolarization tail displays approximately linear behaviour, which is in fact a signature of a power-law dependence of probability. Similar power-law behaviour has been reported in the column density, and has been interpreted as a signature of gravitationally dominated regions undergoing collapse \citep{VSG,BCL,BSC}. We emphasize that we do not know whether self-gravity plays the same role here in the polarization fraction, or rather that some other as yet undetermined mechanism is responsible. We plan to explore possible mechanisms in future work. 

\section{The Dispersion in Polarization Angles} \label{section:S}

\begin{figure*}
	\includegraphics[width=2.1\columnwidth]{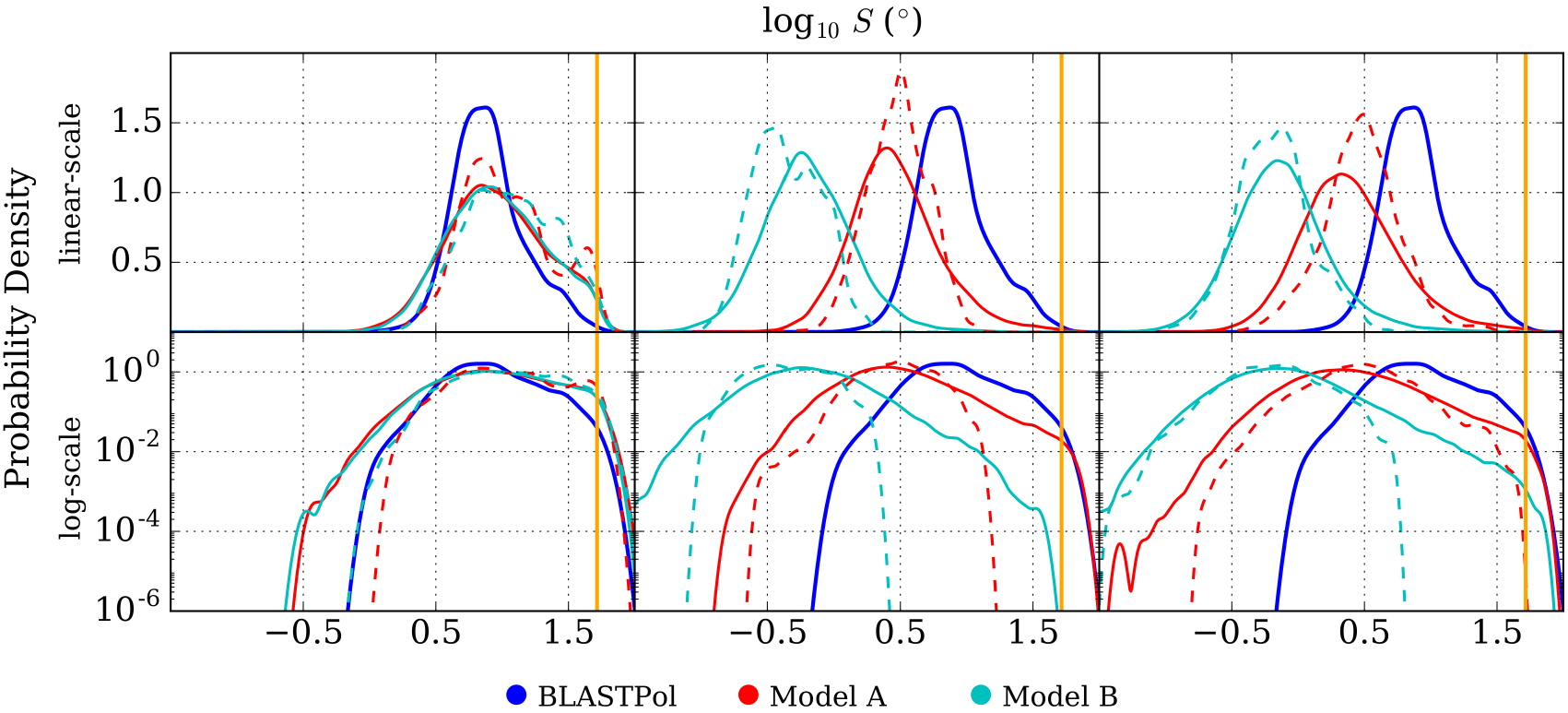}
    \caption{PDFs of the dispersion in polarization angles for the BLASTPol Vela C observations \citep{FBP} (Blue), Model A (Red), and Model B (Cyan). The solid coloured lines correspond to distributions computed at the pixel scale of the simulation, while dashed coloured lines correspond to those observed with a Gaussian beam. The orange line annotates the special value $\pi/\sqrt{12}$. The $x$ line-of-sight distributions are in the left column; in the centre, the $y$ line-of-sight distributions; and in the right, the $z$ line-of-sight distributions. The top row PDFs are log-linear and the bottom row are log-log.}
    \label{fig:SPDF}
\end{figure*}

\begin{table}
	\centering
	\caption{Distribution statistics for the one-dimensional PDFs of the dispersion in polarization angles.}
	\label{tab:SStats}
	\begin{tabular}{cccccc} %
		\hline
		 LOS & Model & Beam & $\mu_G(S)$ & $\log\sigma_G(S)$ & $\overline{\text{Kurt}}_G(S)$ \\
		\hline
		$x$-LOS & A & Pixel   & 9.141$^{\circ}$ & 0.372 & -0.484 \\
                &   & 0.5 pc  & 11.01$^{\circ}$ & 0.338 & -0.654 \\
		        & B & Pixel   & 9.204$^{\circ}$ & 0.364 & -0.516 \\
                &   & 0.5 pc  & 11.51$^{\circ}$ & 0.341 & -0.753 \\
        $y$-LOS & A & Pixel   & 2.904$^{\circ}$ & 0.341 &  0.759 \\
                &   & 0.5 pc  & 3.006$^{\circ}$ & 0.229 & -0.100 \\
		        & B & Pixel   & 0.635$^{\circ}$ & 0.337 &  0.660 \\
                &   & 0.5 pc  & 0.444$^{\circ}$ & 0.251 & -0.724 \\
        $z$-LOS & A & Pixel   & 2.884$^{\circ}$ & 0.379 &  0.460 \\
                &   & 0.5 pc  & 3.327$^{\circ}$ & 0.295 &  0.733 \\
		        & B & Pixel   & 0.746$^{\circ}$ & 0.358 &  1.179 \\  
                &   & 0.5 pc  & 0.635$^{\circ}$ & 0.283 &  0.148 \\
        \textbf{BLASTPol} & - & 0.5 pc & \textbf{7.933$^{\circ}$} & \textbf{0.266} & \textbf{0.222} \\
		\hline
	\end{tabular}
\end{table}

The dispersion in polarization angles is a direct measure of plane-of-sky magnetic disorder. At the smallest scales it measures local changes in magnetic field orientation: high $S$ indicates significant differences in magnetic field behaviour between adjacent sightlines, whereas low $S$ indicates a uniform field that changes little between adjacent sightlines. Rapid changes may be due to dominant dense regions changing magnetic orientation, or due to uniform changes across the whole sightline. Due to the fact that $S$ cannot be scaled (see Section \ref{section:scaling}), determining the mean angle dispersion provides a crucial general measure of plane-of-sky magnetic disorder, which is intrinsic to the MC under observation. Furthermore, the reduced sensitivity of $S$ to heterogeneous grain alignment effects (to be discussed in more detail in a follow-up paper) renders it an even more important measure of magnetic disorder, being strongly tied to the magnetohydrodynamical behaviour of the MC rather than grain alignment microphysics. Different mean angle dispersions point to vastly different conditions due to the angular nature of the quantity, bounded above by 90$^{\circ}$: a mean angle dispersion of 1$^{\circ}$ indicates a very ordered plane-of-sky magnetic field, yet a mean angle dispersion of 10$^{\circ}$ indicates a far higher degree of magnetic disorder in the plane-of-sky. 

Much like what we found in the distributions in polarization fraction, we can see in the dispersion in polarization angles PDFs in Figure \ref{fig:SPDF} that the $x$ line-of-sight (left panel) provides the best agreement with the BLASTPol observations of Vela C. Mean angle dispersions, distribution widths, and kurtosis can be found in Table \ref{tab:SStats}. In general, all simulation distributions have approximately the same width, and are all mesokurtic. But only the mean angle dispersions (of both simulations) in the $x$ line-of-sight are within a degree or so of the BLASTPol value; whereas in the $y$ and $z$ lines-of-sight, the difference is a reduction by a factor of 3 (for the Model A) up to a factor of 10 (for Model B) - significant reductions in the angle dispersion relative to the BLASTPol observations. 

We do find that there is one primary significant difference between Models A and B in the $y$ and $z$ lines-of-sight: the mean angle dispersion is much higher in Model A than Model B. Besides this, the distribution shape is very similar: the widths and kurtosis are nearly the same. As lower mean angle dispersion indicates stronger magnetic order in the plane-of-sky, this indicates that, just as in the case with polarization fraction, higher levels of turbulence and lower magnetization weaken signatures of strong magnetic order. It is also evident that the dispersion in polarization angles shares another common behaviour with the polarization fraction: the $y$ and $z$ lines-of-sight share very similar behaviour for both simulations. In both models, the distributions change little when viewed edge-on ($y$ line-of-sight) or face-on ($z$ line-of-sight), with the dominant magnetic field perpendicular to the line-of-sight. Again, the key difference between these lines-of-sight are the general length of the dense post-shock region along a given sightline, rather than different magnetic field orientation with respect to the plane-of-sky.

This result demonstrates that, in common with the polarization fraction, the dense layer column length has a relatively minor role in not one but both polarimetric observables. It is remarkable that this holds for the dispersion in polarization angles, which is determined by relative changes in the polarization angles on the plane-of-sky. While the polarization angle is determined non-linearly from contributions along the sightline, it can be expected that, in the presence of strong magnetic order in the plane-of-sky, the angles will be dominated by that order (see, e.g., Figure \ref{fig:simpics}, top row). Variations with respect to the mean magnetic field produce structures in the map of the dispersion in polarization angles. As argued before, longer column lengths provide more opportunities to encounter fluctuations with respect to the mean magnetic field, but these fluctuations are hard to interpret with respect to the polarization angle, as these contributions are difficult to gauge in relative importance. On the one hand, higher fluctuation rates might translate to some modest angular differences between the measured plane-of-sky polarization angle and the dominant mean magnetic field. Without some other effect biasing these differences in some direction, they would be expected to be relatively random, and therefore the mean angle dispersion would be relatively higher than measured for a shorter column. On the other hand, angular fluctuations are not purely additive, and any contribution to the line-of-sight averaged quantity would likely be small, and so overall the effect might be negligible. Quantitatively, there appears to be little measurable difference between the $y$ and $z$ line-of-sight views for either simulation (Table \ref{tab:SStats}), so it appears that any effect from the longer column lengths is not important. 
 
The $x$ line-of-sight instead displays a very high mean angle dispersion, centred near the BLASTPol value.  Remarkably, and in common with the polarization fraction, there is little variation between the two models, recovering even the finer structures in the distribution, such as the feature near $\pi/\sqrt{12}$. Note also that no scaling has been applied to the model $S$ distributions. The fact that the two simulations are almost indistinguishable in either polarization fraction contrast or dispersion in polarization angles - two separate measures of magnetic disorder in the plane-of-sky - demonstrates that relative differences in turbulence or magnetization are apparently not very important when polarized emission contributions from the mean magnetic field are suppressed. The differences between this case and the $y$ and $z$ lines-of-sight highlight the fact that apparent magnetic disorder is not a perfect proxy for intrinsic, 3D magnetic disorder; in other words, one cannot conclude that an MC is super-Alfv\'{e}nic from these observations alone. 

Beam convolution affects the dispersion in polarization angles slightly differently than the polarization fraction: not only is resolution degraded, but the lag $\delta$ (see Equation \eqref{S} in Section \ref{section:disp}) is adjusted to reflect the minimum sensible resolution for computation of this quantity. We nevertheless recover similar behaviour that was seen in the beam-convolved polarization fraction, in that beam convolution narrows the width of the distribution, tending to eliminate contributions from both the high and the low dispersion tails. Prior to beam convolution, all three lines-of-sight and both models have very similar distribution widths; after beam convolution, these widths are reduced to similar values as well, and to a value consistent with the width of the BLASTPol distribution. The consistency of the width of the dispersion in polarization angles distributions is a remarkable feature. Taken at face value, this indicates that the width of this distribution is affected neither by the prevailing plane-of-sky magnetic order; the typical column length, or equivalently, cloud depth; nor the level of magnetization/turbulence. Additionally this value is the geometric width, or width in logarithmic space: it appears that values of the dispersion in polarization angles tend to be found within a narrow range of dex from the mean angle dispersion. This could be some indication of a universal property of the dispersion in polarization angles, which merits further investigation, though we emphasize that this is a very tentative conclusion.

We conclude this section by noting the features near the special value $\pi/\sqrt{12}$ (black dashed lines in Figure \ref{fig:SPDF}). As noted above, this is the value to which a collection of random vectors converges. Without exception, all the dispersion in polarization angles distributions rapidly fall off above this value. This is consistent with our expectation, as any value higher than $\pi/\sqrt{12}$ is a very unlikely configuration on the plane-of-sky. This special value also happens to be near the relatively high values of the filamentary-type structures that can be seen in the maps of $S$, found in the bottom panels of Figures \ref{fig:simpics} and \ref{fig:simpicsBM}. These features have been reported in past observations of polarized submillimeter continuum by both Planck \citep{PXIX} and BLASTPol \citep{FBP}. We plan to explore the nature of these filamentary features in future work. 

\section{The Polarimetric Joint Correlation} \label{section:Sp}

\begin{figure*}
	\centering
	\includegraphics[width=2.1\columnwidth]{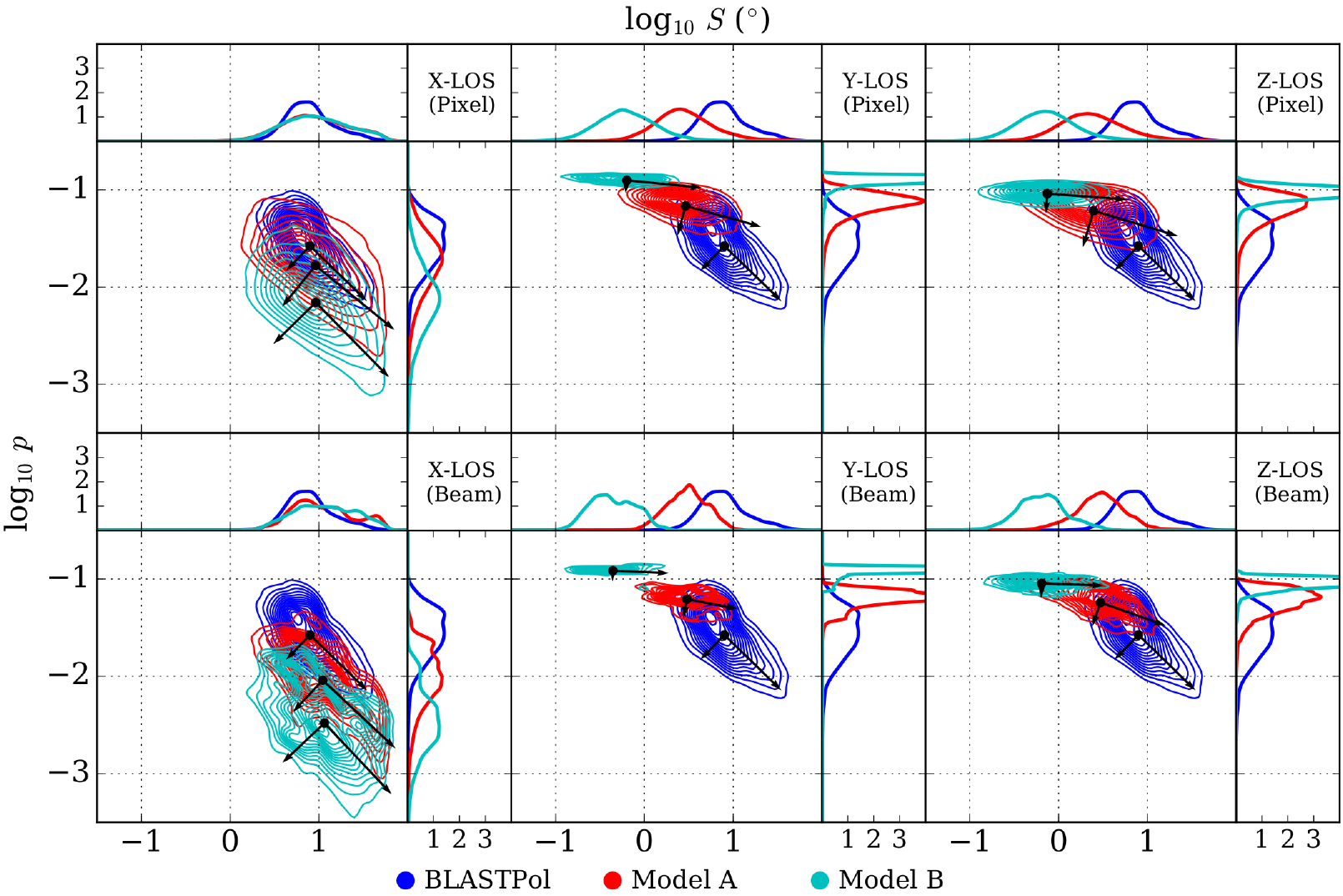}
	\caption{Joint PDFs of the dispersion in polarization angles and the polarization fraction for the BLASTPol Vela C observations \citep{FBP} (Blue), Model A (Red), and Model B (Cyan). Top row is at pixel resolution and the bottom row is convolved with a Gaussian beam. Annotated are the 1D PDFs of the two distributions. The principal component vectors are annotated by vectors; the principal eigenvalues, eigenvalue ratios, and implied power-law indices are given in Table \ref{tab:SpPCA}.}
    \label{fig:SP}
\end{figure*}

\begin{table}
\centering
	\caption{Principal Component Analysis implied power-law index and Pearson/Spearman correlation coefficients for the dispersion in polarization angles versus polarization fraction joint PDF.}
	\label{tab:SpPCA}
	\begin{tabular}{cccccc} %
		\hline
		 LOS & Model & Beam & Index & $\rho_{P,Sp}$ & $\rho_{S,Sp}$ \\
		\hline
		$x$-LOS & A & Pixel   & -0.814 & -0.577 & -0.555 \\
                &   & 0.5 pc  & -0.936 & -0.682 & -0.660 \\
		        & B & Pixel   & -1.020 & -0.514 & -0.555 \\
                &   & 0.5 pc  & -1.058 & -0.479 & -0.494 \\
        $y$-LOS & A & Pixel   & -0.275 & -0.529 & -0.495 \\
                &   & 0.5 pc  & -0.190 & -0.439 & -0.413 \\
		        & B & Pixel   & -0.098 & -0.507 & -0.385 \\
                &   & 0.5 pc  & -0.043 & -0.234 & -0.008 \\
        $z$-LOS & A & Pixel   & -0.321 & -0.511 & -0.419 \\
                &   & 0.5 pc  & -0.385 & -0.645 & -0.600 \\
		        & B & Pixel   & -0.072 & -0.292 & -0.109 \\  
                &   & 0.5 pc  & -0.037 & -0.165 & -0.163 \\
        \textbf{BLASTPol} & - & 0.5 pc & \textbf{-0.969} & \textbf{-0.684} & \textbf{-0.635} \\
		\hline
	\end{tabular}
\end{table}

Thus far, our findings present consistent behaviour for both the polarization fraction and the dispersion in polarization angles. The BLASTPol observations are well-matched by the $x$ line-of-sight distributions of each respectively; the distributions for the $y$ and $z$ lines-of-sight, while very consistent with each other, provide a poor match. These conclusions are supported by the distributions of the observables alone, as noted in Sections \ref{section:p} and \ref{section:S}, but we can carry the analysis further and ask whether they agree with respect to the joint correlation of the two observables. This joint correlation has been studied by both BLASTPol \citep{FBP} and Planck \citep{PXIX,PXX} in the past. For brevity, we will refer to this particular joint correlation between $p$ and $S$ as the polarimetric joint correlation. The joint correlation contains strictly more information than the distributions of each observable alone, which may be thought of as projections of the joint distribution onto a single axis. Any apparent agreement between the $x$ line-of-sight and the BLASTPol observations is substantially weakened if the joint correlation behaviour of the two are not consistent. 

The polarimetric joint correlations can be found in Figure \ref{fig:SP}, both at the pixel resolution (top panel) and after beam convolution (bottom panel). The implied power-law indices derived from the principal components, as well as the Pearson and Spearman correlation coefficients, can be found in Table \ref{tab:SpPCA}. For clarity, we present the joint correlation computed without applying a scaling to the polarization fraction, which would obscure the contours. (The reader may apply this scaling visually by moving the distribution along the $p$ axis of the plot.) It is evident that the $x$ line-of-sight provides excellent agreement with the BLASTPol observations: the power-law indices are very close to each other (being close to -1) and the correlation coefficients indicate a moderately strong correlation on the same order as the BLASTPol data. In contrast, neither of the simulations in the $y$ and $z$ lines-of-sight provide very strong agreement, with power-law indices generally significantly shallower than the BLASTPol index. The correlations can be both weaker than the BLASTPol correlation (e.g. Model B in the $z$ line-of-sight) or around the same order as the BLASTPol result (e.g. Model A in the $x$ line-of-sight). Additionally, as demonstrated in Section \ref{section:S}, the mean angle dispersion is too low in the $y$ and $z$ lines-of-sight, which reduces the agreement further, but we also note that this is not new information as revealed by the joint correlation. Beam convolution does not appear to modify the joint correlations substantially; the differences between the lines-of-sight remain significantly more important. 

Both simulations when viewed from all three lines-of-sight, and the BLASTPol observations, consistently demonstrate a negative power-law index for the polarimetric joint correlation; this has been noted in other work as well \citep{PXIX}. This negative dependence is expected between polarization fraction and dispersion in polarization angles. On the one hand, highly depolarized sightlines may be regions with high inclination with respect to the line-of-sight, and these regions could have a high dispersion in polarization angles if this inclination is significantly different from the mean magnetic field in the region. On the other hand, sightlines might contain a significant degree of cancellation (resulting in depolarization) if there are significant changes in the magnetic field orientation in that region; if these changes are at all in the plane-of-sky direction, then they would show up in the dispersion in polarization angles, raising it. However, the correlation is certainly far from perfect. For each polarization fraction there is a fairly wide range in dispersion in polarization angles, and vice versa. Moreover, any power-law index calculation, be it based on the principal components or based on linear regression, will be dominated by the highest probability density regions. Given the very different asymptotic behaviour of the polarization fraction and dispersion in polarization angles, the power-law index likely encapsulates the behaviour of the most common regions found between the extremes in either quantity. 

In Sections \ref{section:p} and \ref{section:S}, we noted the common behaviour in the polarization fraction and dispersion in polarization angles distributions for each simulation when viewed from the $y$ and $z$ lines-of-sight, which indicates the insensitivity of these distributions to the typical column length. We find this behaviour again in the polarimetric joint correlation, in this case in terms of the power-law index and correlation coefficients. The shallowness of the power-law index appears to be another signature of high plane-of-sky magnetic order. Interpreting this in terms of the relation between magnetic disorder as measured by $S$ and magnetic disorder as measured by $p$, then the shallow power-law is likely a consequence of the significantly reduced width in the polarization fraction. This offers a clue into what kind of magnetic disorder each polarimetric observable measures: the polarization fraction is dependent on both inclination and cancellation within the line-of-sight. A strong plane-of-sky magnetic field tends to directly reduce inclination. When projecting the small variations from the mean-magnetic field onto the plane-of-sky, the angular differences should nevertheless remain small. When the mean-magnetic field is instead parallel to the line-of-sight, the magnetic field tends to produce a high degree of inclination within the sightline. Additionally, small variations with respect to the mean magnetic field, even though they might be small, could introduce significant angular differences within the column once projected onto the plane-of-sky, raising the effects of cancellation. We have also noted in Section \ref{section:S} that the width of the dispersion in polarization angles distribution changes very little between different lines-of-sight or between Model A and Model B; the distribution is instead set by the mean angle dispersion, which defines the overall level of plane-of-sky magnetic disorder. In the presence of strong plane-of-sky magnetic order (the $y$ and $z$ lines-of-sight) $S$ is only weakly anti-correlated with $p$, whereas a strong correlation is noted when the plane-of-sky mean magnetic field is suppressed (the $x$ line-of-sight). Therefore, strong plane-of-sky magnetic order de-correlates these two measures of magnetic disorder. 

The one-dimensional distributions of polarization fraction and dispersion in polarization angles alone suggest that the plane-of-sky magnetic field in Vela C is inconsistent with an ordered plane-of-sky magnetic field such as that found in both Models when viewed from the $y$ and $z$ lines-of-sight. Given that the agreement persists not only in distribution shape and extent but also mutual dependence, this suggestion is strengthened significantly. This can be interpreted in several ways. On the one hand, the intrinsic magnetic disorder of the three lines-of-sight do not change for any given simulation; the only change is the angle at which it is viewed. It may be that Vela C happens to be serendipitously aligned relative to our viewing angle, and that contributions from the highly ordered mean magnetic field are suppressed in Vela C. However, it is very unlikely for Vela C to have the highly ordered mean magnetic field of our highly idealized simulations. 

On the other hand, we note that the three key quantities indicating agreement/disagreement - polarization fraction distribution width, mean angle dispersion, and polarimetric correlation power-law index - differ between models A and B in the $y$ and $z$ lines-of-sight, indicating dependence on the level of turbulence and magnetization in the presence of strong magnetic order in the plane-of-sky. A higher turbulence level/lower magnetization (model A) has better agreement (with respect to these three quantities) with the BLASTPol values. It may be possible that agreement can be achieved by increasing the Alfv\'{e}n Mach number in the post-shock region (thus increasing the importance of turbulence and decreasing the importance of magnetization). The sensitivity of these key quantities on turbulence and magnetization, in this context, supports the notion that the $x$ line-of-sight distributions (and the observed BLASTPol distributions) are realizations of an apparently highly disordered plane-of-sky magnetic field. If this is the case, then there is a degeneracy: it may not be possible to distinguish between the effects of high super-Alfv\'{e}nic turbulence and suppression of the mean magnetic field in the plane-of-sky purely through studying the polarimetric distributions alone. It may be possible to break this degeneracy though the use of velocity information obtained from molecular line observations.

\section{Joint Correlations Involving Column Density} \label{section:N}

\begin{figure*}
    \centering
	\includegraphics[width=2.1\columnwidth]{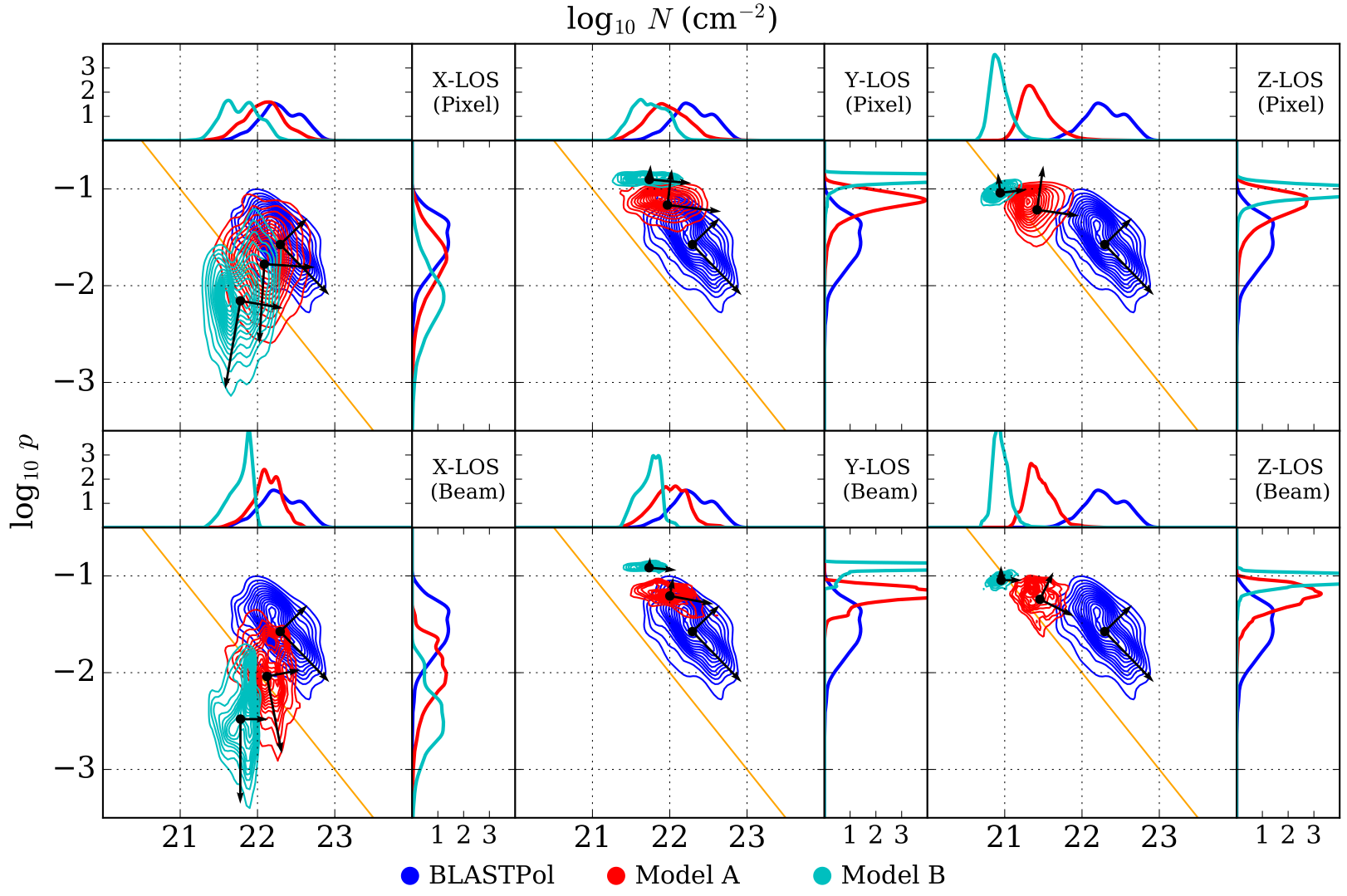}
	\caption{Joint PDFs of the column density and the polarization fraction for the BLASTPol Vela C observations \citep{FBP} (Blue), Model A (Red), and Model B (Cyan). Top row is at pixel resolution and the bottom row is convolved with a Gaussian beam. Annotated are the 1D PDFs of the two distributions. The principal component vectors are annotated by vectors; the principal eigenvalues, eigenvalue ratios, and implied power-law indices are given in Table \ref{tab:IpPCA}. The annotated orange lines represent the polarization sensitivity limit to BLASTPol, as discussed in Section \ref{section:N}.}
    \label{fig:IP}
\end{figure*}

\begin{table}
\centering
	\caption{Principal Component Analysis implied power-law index and Pearson/Spearman correlation coefficients for the column density versus polarization fraction joint PDF.}
	\label{tab:IpPCA}
	\begin{tabular}{cccccc} %
		\hline
		 LOS & Model & Beam & Index & $\rho_{P,Np}$ & $\rho_{S,Np}$ \\
		\hline
		$x$-LOS & A & Pixel   &  15.59 &  0.029 & 0.026 \\
                &   & 0.5 pc  & -5.168 & -0.240 & -0.227 \\
		        & B & Pixel   &  5.794 &  0.174 & 0.171 \\
                &   & 0.5 pc  &  333.3 &  0.005 &  0.036 \\
        $y$-LOS & A & Pixel   & -0.121 & -0.156 & -0.137 \\
                &   & 0.5 pc  & -0.184 & -0.412 & -0.366  \\
		        & B & Pixel   & -0.088 & -0.276 & -0.073 \\
                &   & 0.5 pc  & -0.091 & -0.258 & -0.045  \\
        $z$-LOS & A & Pixel   & -0.064 & -0.026 & 0.036 \\
                &   & 0.5 pc  & -0.417 & -0.252 & -0.198  \\
		        & B & Pixel   &  0.117 &  0.122 & 0.436 \\  
                &   & 0.5 pc  &  0.009 &  0.011 &  0.327  \\
        \textbf{BLASTPol} & - & 0.5pc & \textbf{-1.029} & \textbf{-0.542} & \textbf{-0.564} \\
		\hline
	\end{tabular}
\end{table}

\begin{figure*}
	\centering
	\includegraphics[width=2.1\columnwidth]{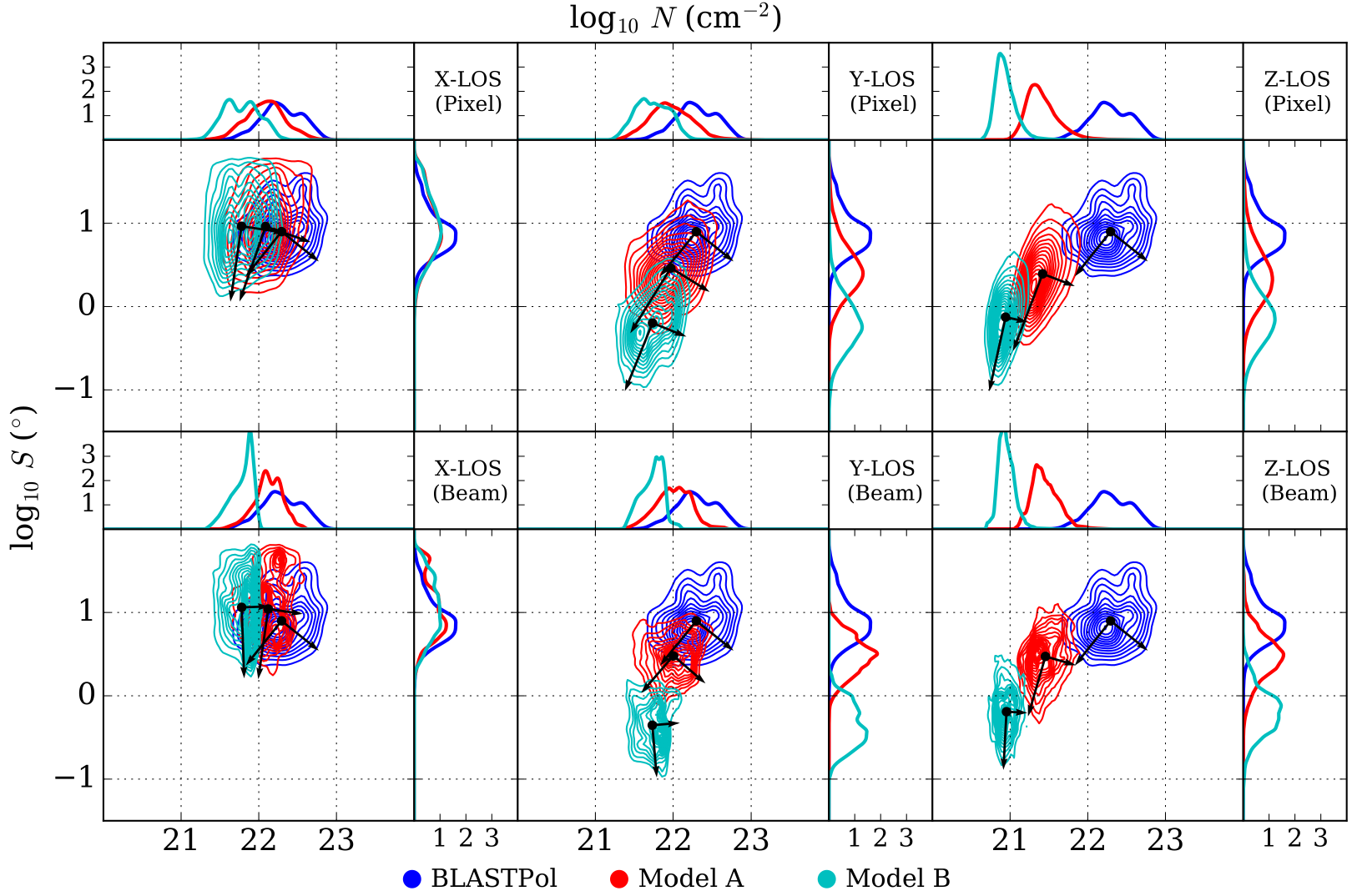}
	\caption{Joint PDFs of the column density and the dispersion in polarization angles for the BLASTPol Vela C observations \citep{FBP} (Blue), Model A (Red), and Model B (Cyan). Top row is at pixel resolution and the bottom row is convolved with a Gaussian beam. Annotated are the 1D PDFs of the two distributions. The principal component vectors are annotated; the principal eigenvalues, eigenvalue ratios, and implied power-law indices are given in Table \ref{tab:ISPCA}.}
    \label{fig:IS}
\end{figure*}

\begin{table}
\centering
	\caption{Principal Component Analysis implied power-law index and Pearson/Spearman correlation coefficients for the column density versus dispersion in polarization angles joint PDF.}
	\label{tab:ISPCA}
	\begin{tabular}{cccccc} %
		\hline
		 LOS & Model & Beam & Index & $\rho_{P,NS}$ & $\rho_{S,NS}$ \\
		\hline
		$x$-LOS & A & Pixel   &  2.881 &  0.280 &  0.259 \\
                &   & 0.5 pc  &  7.021 &  0.189 &  0.172 \\
		        & B & Pixel   &  6.728 &  0.145 &  0.148 \\
                &   & 0.5 pc  &  28.54 & -0.069 & -0.089 \\
        $y$-LOS & A & Pixel   &  1.613 &  0.421 &  0.366 \\
                &   & 0.5 pc  &  1.159 &  0.140 &  0.153 \\
		        & B & Pixel   &  2.456 &  0.445 &  0.422 \\
                &   & 0.5 pc  & -11.72 & -0.105 & -0.111 \\
        $z$-LOS & A & Pixel   &  2.382 &  0.493 &  0.472 \\
                &   & 0.5 pc  &  3.350 &  0.282 &  0.283 \\
		        & B & Pixel   &  4.544 &  0.510 &  0.441 \\  
                &   & 0.5 pc  &  18.85 &  0.120 &  0.098 \\ 
        \textbf{BLASTPol} & - & 0.5 pc & \textbf{1.250} & \textbf{0.164} & \textbf{0.167} \\
		\hline
	\end{tabular}
\end{table}

We are not limited to considering only the polarization fraction and dispersion in polarization angles in our comparisons. Correlations involving the column density have long been studied as part of both observational polarimetric studies \citep{WLH,PXIX,FBP} and theoretical investigation \citep{CL1,FG1,PFG,PXX,CKL}. Much work has focused on the polarization fraction - column density joint correlation, which has typically been fit to a form $p \propto N^{\gamma}$, where $\gamma$ is a power-law index called the depolarization parameter in \citet{PFG}. The depolarization parameter, and its equivalent for the dispersion in polarization angles - column density joint correlation, has usually been found by computing linear regression in log-log space \citep{FBP}. Using our alternative measure using the principal components, we derive a depolarization parameter for BLASTPol equal to -1.029, which is about twice that quoted in \citet{FBP}; this dependence is moderately strong as measured by the correlation coefficients (see Table \ref{tab:IpPCA}). This is consistent with other quoted values in the literature which range from -0.5 to -1.5 \citep{CL1,FG1,PFG,PXX,CKL}. The dispersion in polarization angles - column density joint correlation has been less well explored. For BLASTPol, we determine a power-law index of 1.250, which is much higher than the value reported in \citet{FBP}; nonetheless, we also find low correlation coefficients, which indicate a weak dependence (see Table \ref{tab:ISPCA}). 

We first turn our attention to the polarization fraction - column density joint correlations, which may be found in Figure \ref{fig:IP}. The depolarization parameters and correlation coefficients may be found in Table \ref{tab:IpPCA}. We note first that neither the $x$ line-of-sight nor the $y$ and $z$ lines-of-sight agree very well with the BLASTPol data. There is similar behaviour between the $y$ and $z$ lines-of-sight, as we might expect given the similarity of the polarization fraction distributions for these two lines-of-sight. The depolarization parameter for these lines-of-sight is generally weakly negative in most cases (with the exception of the $z$ line-of-sight view of Model B) and the correlations are generally poor ($|\rho_P|$ and $|\rho_S| << 1$). On the other hand, the $x$ line-of-sight displays significantly different behaviour, and has depolarization parameters deviating far from the normally quoted values ($-1.5 < \gamma < -0.5$) for both models. The very low correlation coefficients indicate that a power-law dependence may not necessarily be descriptive for these distributions. In all cases, beam convolution distorts the joint distributions, but generally only introduces modest changes to the principal components.\footnote{In the case of the $x$ line-of-sight, the principal component points nearly straight down where there is an infinite discontinuity. It changes from negative to positive infinity as the principal component rotates counter-clockwise through this point.}

The column density-polarization fraction correlation might be disproportionately affected by unavoidable limitations to the BLASTPol polarization sensitivity. BLASTPol measurements require a sufficient signal-to-noise measurement of the polarized intensity, which is proportional to the product of the polarization fraction and column density. This limitation would eliminate portions of the joint correlation that could not produce a sufficiently high polarized intensity. However, the estimated minimum polarization fraction that BLASTPol could measure is $10^{-3}$ \citep{FBP}; thus the minimum polarized intensity BLASTPol requires (in column density units) is on the order of $10^{20}$ cm$^{-2}$.\footnote{This can be derived from the aforementioned minimum polarization fraction quoted in \citet{FBP} and the maximum column densities in those same observations.} An orange line annotating this threshold is included in the plots in Figure \ref{fig:IP}. There is some distance between the edge of the BLASTPol distribution and this line; while some small portion of the distribution may have been lost, it is very unlikely that enough sightlines were lost to significantly alter the principal components, which are dominated by a large number of measurements well within the range of BLASTPol.\footnote{This sensitivity limit does not necessarily apply to the synthetic observations; we can shift the distributions above the sensitivity limit using an appropriate scaling transformation (see Section \ref{section:scaling}) and choice of $p_0$, which we have not done for clarity, to avoid crowding the contours in the plot.} We do not believe this effect can explain any lack of agreement between our simulations and the BLASTPol observations.  

Next, we consider the dispersion in polarization angles - column density joint correlations, found in Figure \ref{fig:IS}. The principal component-implied power-law indices for these correlations may be found in Table \ref{tab:ISPCA}, along with the correlation coefficients $\rho_P$ and $\rho_S$. The $z$ line-of-sight displays moderately stronger correlations than BLASTPol or the other lines-of-sight; Model A has moderately positive correlations, while those of Model B are very steep. This is possibly just a reflection of the relative width of the column density distributions in this line-of-sight when compared to the width of the $S$ distribution; in the other lines-of-sight the widths are relatively comparable, but in the $z$ line-of-sight the column density width is narrower. In general, the rest of the joint correlations display weak dependence, and beam convolution tends to reduce the strength of these correlations. Altogether, our results tend to agree with the conclusion that the dispersion in polarization angles is relatively uncorrelated with column density, as was reported in BLASTPol.

Correlations with the column density have been interpreted in multiple ways. The depolarization parameter has commonly been used to constrain the microphysics of grain alignment, which may vary due to local conditions in the cloud \citep{CL1,WHLH}. Consistent with previous efforts \citep{FG1} we find a limited amount of depolarization that arises purely from magnetohydrodynamical effects, but these effects are weak if the principal component derived slopes are to be believed. The shape of the joint correlations and their weak correlation coefficients suggest that these effects are weaker still, and given the inability of our simulations to achieve agreement with the BLASTPol results - or any previously reported depolarization parameters - it is likely that the assumption of homogeneous grain alignment could be responsible. Heterogeneous grain alignment, where polarization efficiency depends on local conditions such as the gas density, may affect the results we have presented. Corrections to the Stokes parameters would be present in first order in the polarization fraction; however, the polarization angle (as it is computed from the arctangent of a ratio) is less sensitive to these corrections. Therefore, the dispersion in polarization angles is likely less sensitive to these effects. We will address the role of heterogeneous grain alignment, and its effect on the polarimetric observables, in subsequent work. 

\section{Intermediate Inclination} \label{section:inc}

\begin{figure*}
	\includegraphics[width=2.1\columnwidth]{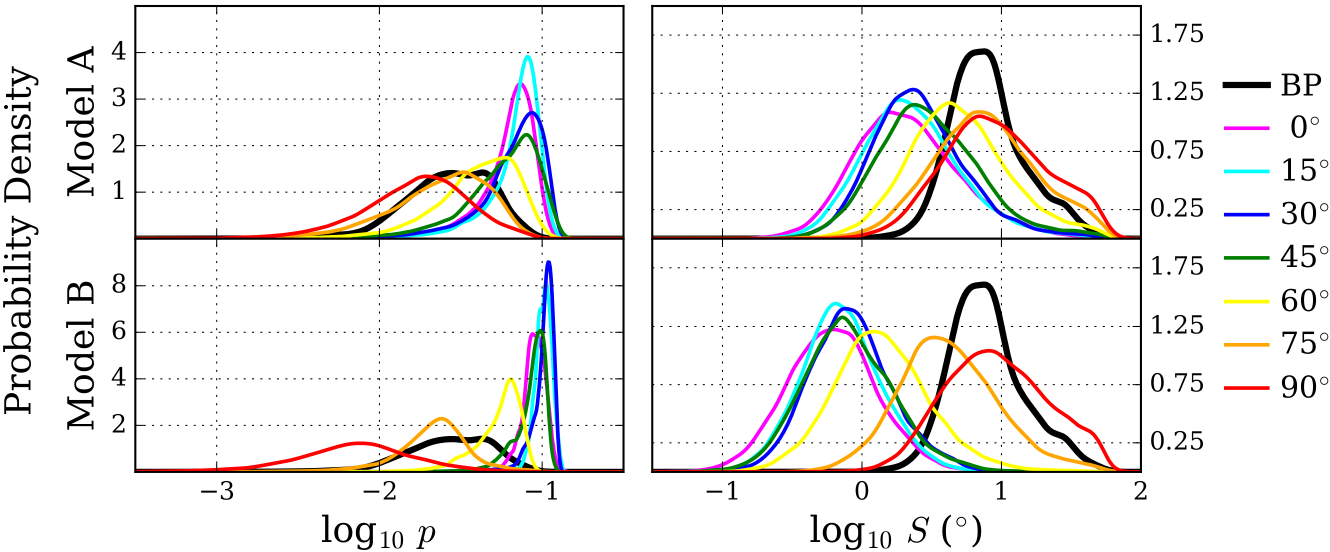}
	\caption{PDFs of the polarization fraction (left column) and dispersion in polarization angles (right column) for both Model A (top row), and Model B (bottom row) after rotating by $0^{\circ}$ (violet, identical to the $z$ line-of-sight); $15^{\circ}$ (cyan); $30^{\circ}$ (blue); $45^{\circ}$ (green); $60^{\circ}$ (yellow); $75^{\circ}$ (orange); and $90^{\circ}$ (red, identical to the $x$ line-of-sight). The BLASTPol distribution is annotated in black.}
    \label{fig:inclination}
\end{figure*}

\begin{figure*}
	\includegraphics[width=2.1\columnwidth]{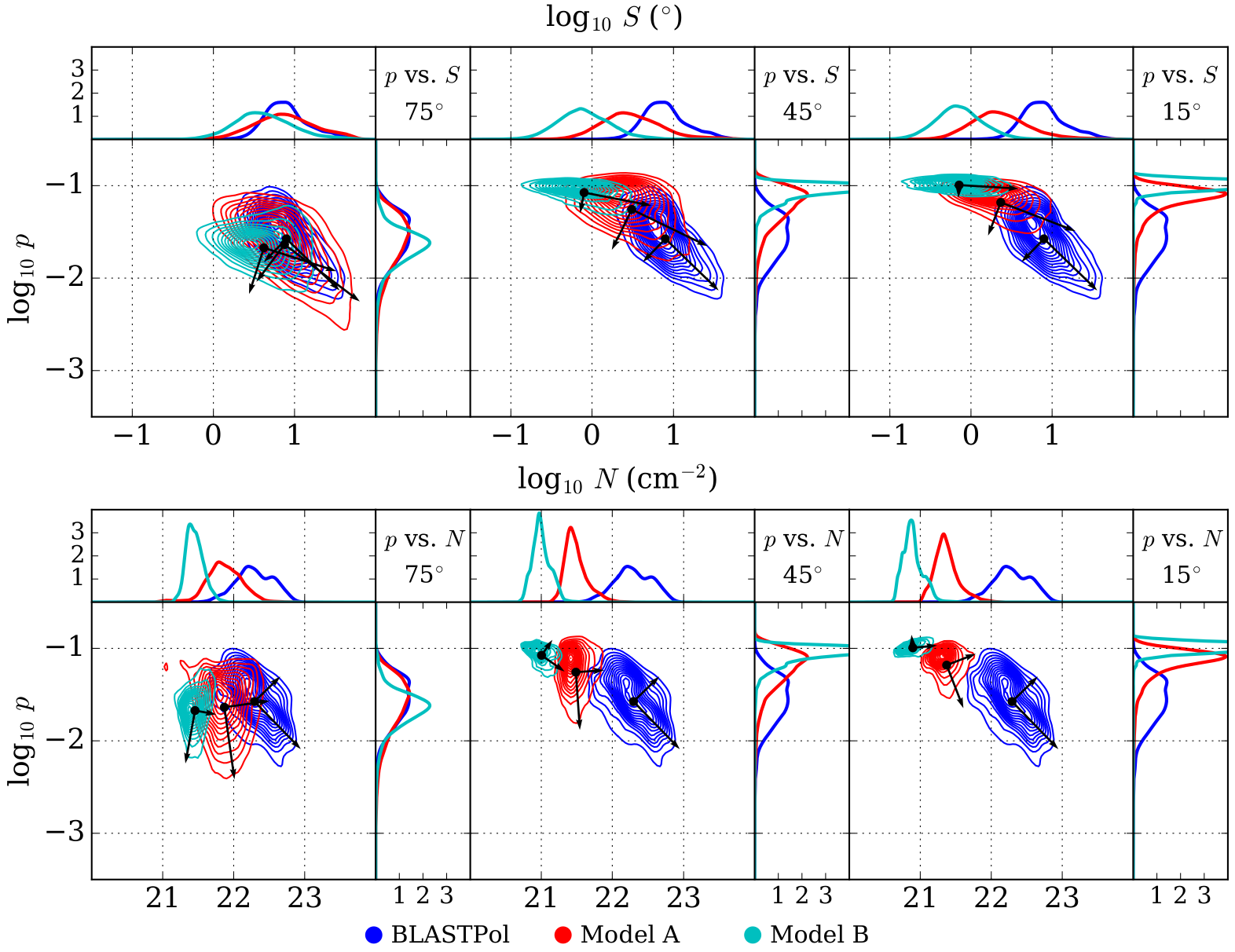}
	\caption{Joint PDFs of the dispersion in polarization angles vs. polarization fraction (top row) and column density vs. polarization fraction (bottom row) for three inclination angles, intermediate between 0$^{\circ}$ (magnetically ordered, face-on view, $z$ line-of-sight) and 90$^{\circ}$ (magnetically disordered, edge-on view, $x$ line-of-sight): 15$^{\circ}$ (right column), 45$^{\circ}$ (centre column), and 75$^{\circ}$ (left column).}
    \label{fig:inclination2}
\end{figure*}

\begin{table*}
\centering
	\caption{Mean polarization fraction, mean dispersion in polarization angles, and polarization fraction distribution widths for the $p$ and $S$ distributions; and power law indices inferred from the principal components, and the Pearson and Spearman rank correlation coefficients, for the $S$ vs $p$ and $N$ vs $p$ joint correlations, all computed for several inclination angles between the $z$ and $x$ lines-of-sight.}
    \label{tab:inclination}
	\begin{tabular}{ccccccccccc}
    	\hline
        Angle & Model & $\mu_G(p)$ & $\mu_G(S)$ & $\log \sigma_G(p)$ & $S-p$ & $\rho_{P,Sp}$ & $\rho_{S,Sp}$ & $N-p$ & $\rho_{P,Np}$ & $\rho_{S,Np}$ \\
        	  &       &            &            &                    & Index &               &               & Index &               &               \\
        \hline
		15$^{\circ}$ & A & 0.066 & 2.33$^{\circ}$ & 0.201 & -0.383 & -0.571 & -0.467 & -2.567 & -0.132 & -0.083\\
                     & B & 0.101 & 0.71$^{\circ}$ & 0.058 & -0.060 & -0.292 & -0.206 & ~0.099 & ~0.193 & ~0.356\\
		30$^{\circ}$ & A & 0.064 & 2.61$^{\circ}$ & 0.225 & -0.471 & -0.536 & -0.408 & -4.477 & -0.151 & -0.156\\
                     & B & 0.101 & 0.84$^{\circ}$ & 0.073 & -0.125 & -0.487 & -0.393 & -0.198 & -0.262 & -0.103\\
        45$^{\circ}$ & A & 0.055 & 3.05$^{\circ}$ & 0.250 & -0.483 & -0.517 & -0.391 & -14.46 & -0.065 & -0.028\\
                     & B & 0.084 & 0.80$^{\circ}$ & 0.110 & -0.193 & -0.536 & -0.468 & -0.703 & -0.431 & -0.436\\
        60$^{\circ}$ & A & 0.039 & 4.96$^{\circ}$ & 0.263 & -0.540 & -0.510 & -0.419 & ~4.876 & ~0.124 & ~0.173\\
                     & B & 0.053 & 1.38$^{\circ}$ & 0.159 & -0.274 & -0.520 & -0.441 & -1.889 & -0.367 & -0.408\\
        75$^{\circ}$ & A & 0.023 & 7.53$^{\circ}$ & 0.315 & -0.757 & -0.614 & -0.583 & -7.174 & -0.059 & -0.041\\
                     & B & 0.021 & 4.19$^{\circ}$ & 0.231 & -0.322 & -0.324 & -0.251 & ~5.497 & ~0.245 & ~0.205\\
		\hline
	\end{tabular}
\end{table*}

The excellent agreement with BLASTPol provided by the $x$ line-of-sight, with respect to the distributions of the polarimetric observables and the polarimetric joint correlation, lends weight to the conclusion that Vela C might have common plane-of-sky magnetic structure with the $x$ line-of-sight. Consistently we have demonstrated that it is possible to achieve polarimetric distributions strikingly similar to the BLASTPol distributions provided that the mean magnetic field is not in the plane-of-sky, as is the case in the $x$ line-of-sight. However, this is a very specific arrangement in a highly idealized simulation: a real MC, such as Vela C, will very likely not have an ordered mean magnetic field that happens to be perfectly aligned with respect to the observer. One may ask: does the agreement with BLASTPol requires such an orientation? Is it possible to achieve agreement with the BLASTPol distributions with a mean magnetic field that is only moderately suppressed with respect to the line-of-sight? We explore such an arrangement by computing synthetic observations along lines-of-sight inclined at some angle between the $x$ line-of-sight direction and the $z$ line-of-sight.\footnote{We compare only to the $z$ line-of-sight given the similarity of the $y$ line-of-sight and $z$ line-of-sight distributions demonstrated in Sections \ref{section:p} and \ref{section:S}.} In such an arrangement the mean magnetic field has some component in the plane-of-sky, but its ordering influence is reduced in comparison to the extreme cases. 

Figure \ref{fig:inclination} contains the distributions of polarization fraction and dispersion in polarization angles (computed at pixel scale) for a range of intermediate inclination angles between the $z$ and $x$ lines-of-sight. The mean polarization fraction, mean angle dispersion, and width in polarization fraction are presented in Table \ref{tab:inclination}. The convention is that $0^{\circ}$ indicates the $z$ line-of-sight and $90^{\circ}$ indicates the $x$ line-of-sight. As the inclination angle increases, the polarization fraction distributions gradually increase in width; their peaks become more suppressed, becoming closer to the BLASTPol distribution. Similarly, the mean angle dispersion increases with increasing inclination angle, raising the overall level of the $S$ distribution. The Model A distributions become closer to the BLASTPol distribution at a smaller inclination angle than the Model B distributions do (compare the 60$^{\circ}$ and 75$^{\circ}$ distributions for Model A to Model B in Figure \ref{fig:inclination}), but agreement is reached before completely reaching the $x$ line-of-sight. We find similar behaviour in the polarimetric joint correlation, which is presented for a few intermediate inclination angles in the top row of Figure \ref{fig:inclination2}; the power-law indices and correlation coefficients are also found in Table \ref{tab:inclination}. The distributions steepen (as measured by power-law index) as the inclination angle increases, and become closer to the BLASTPol distribution. Again, Model A reaches agreement with BLASTPol before Model B; Model A has nearly identical principal components to BLASTPol in the 75$^{\circ}$ inclination (left column, Figure \ref{fig:inclination2}), yet Model B remains shallower. Based on these comparisons, it is clear that Model A has consistency with BLASTPol at inclinations greater than at least 75$^{\circ}$ (measured from the $z$ line-of-sight), while Model B would require higher inclinations. 

These results suggest that the BLASTPol observations are consistent not just with one single orientation but a range of possible inclinations. Additionally, we see again that higher levels of turbulence (Model A) are consistent with a wider range of inclination angles; we can see this by directly comparing the Model A and B distributions for each inclination angle in Figure \ref{fig:inclination}. At face value, one might conclude that Vela C shares the same magnetic structure as our simulations (within some range of possible inclinations) which may vary depending on the relative level of turbulence and magnetization. However, we note that our simulations only represent simplified, idealized scenarios. Another, perhaps more likely, interpretation of the inclination angle is as a mixing angle, indicating the proportion of the potentially irregular mean magnetic field that happens to be in the plane-of-sky. This would indicate that the polarimetric distributions in Vela C are at least partially accounted for by suppression of the mean magnetic field to some degree with respect to the plane-of-sky. 

Regardless of the interpretation, our results demonstrate that there is a significant degeneracy in apparent magnetic disorder. The key polarimetric signatures of intrinsic magnetic disorder - high mean angle dispersion, wide distributions in polarization fraction, and a steep negative correlation between these two - can arise either from a highly turbulent and low magnetization environment (intrinsic magnetic disorder) or from suppression of the organizing influence of the mean magnetic field in the plane-of-sky (a projection effect). Each of these signatures in the $y$ and $z$ lines-of-sight are closer to the BLASTPol values in Model A than Model B; and as mentioned, agreement with BLASTPol is reached for Model A at less extreme inclination than Model B. Highly disordered magnetic fields have long been suspected of arising due to high turbulence and/or low magnetization; however the signatures of magnetic disorder can be affected as much or more by reducing the plane-of-sky component of the mean magnetic field, which can, in principle, be relatively strong. The consistency of the BLASTPol data with the $x$ line-of-sight distributions of Model B demonstrate this possibility, in which sub-Alfv\'{e}nic conditions are nevertheless consistent with apparent signatures of high magnetic disorder.

One may also ask if an intermediate inclination angle can address the issues raised in Section \ref{section:N} with respect to the joint correlation between polarization fraction and column density. It is conceivable that an intermediate inclination angle may result in a principal component consistent with BLASTPol and other observations of the depolarization parameter. If that were the case, then a suitable inclination angle choice may result in the appropriate power-law index consistent with BLASTPol, though this might require some fine-tuning. The bottom row of Figure \ref{fig:inclination2} contains the column density-polarization fraction joint correlations for a few inclination angles; the power-law indices may be found also in Table \ref{tab:inclination}. There is a modest agreement with the Model A column density-polarization fraction distribution at 15$^{\circ}$ inclination (right column bottom row, Figure {fig:inclination2}), which is peculiar given the evidence for agreement at large rather than small inclination angles in the polarization fraction and dispersion in polarization angles. It appears that the relative importance of the vertically oriented principal component (dominated by the width of the polarization fraction) increases as inclination angle increases. This indicates that the difference between the $x$ line-of-sight and the $y$ and $z$ lines-of-sight appears to be an expression of the previously identified increase in the width of polarization fraction. Additionally, the discrepancy in correlation strength (as measured by the correlation coefficients in Table \ref{tab:inclination}) does not improve with inclination; they remain significantly low compared to the BLASTPol correlation coefficients in nearly all cases, including $15^{\circ}$ (see Table \ref{tab:IpPCA}). In this light, the modest agreement at $15^{\circ}$ is likely a coincidence. We conclude that inclination cannot explain the disagreement between our synthetic observations and the BLASTPol observations, which supports the notion we articulated earlier that the column density correlations need to be explored in the context of heterogeneous grain alignment. 

\section{Conclusions} \label{section:conc}

We provide a direct comparison between the BLASTPol observations of the Vela C molecular ridge \citep{FBP} and numerical simulations of MCs with colliding flows \citep{CO2}. We perform this direct comparison by computing synthetic polarimetry of two numerical simulations (Models A and B; see Table \ref{tab:simpars}), and applying the same statistical analysis methods to both the synthetic observations and the BLASTPol polarimetry data for the Vela C  molecular cloud. The BLASTPol observations provide an unprecedentedly high number of polarization psuedovectors for a single molecular cloud, enabling our use of detailed statistical comparison with numerical simulations. Our main conclusions are the following:

\begin{enumerate}

\item{We find that the distribution of polarization fraction $p$ in Vela C is rather broad when compared to our simulations observed with the mean magnetic field parallel to the plane-of-sky (Section \ref{section:p}), and shows remarkable consistency with the $x$ line-of-sight, in which the mean magnetic field is mostly parallel to the line-of-sight (Figure \ref{fig:pPDF}). We find that those lines-of-sight in which the mean magnetic field is primarily in the plane-of-sky (the $y$ and $z$ lines-of-sight) produce much more highly peaked polarization fraction distributions, indicating little variability. In this regime, the width of the polarization fraction PDF appears to be related to the level of turbulence and magnetization, and is not substantially affected by differences in the amount of material in the sightlines; higher turbulence/lower magnetization widens the distribution. In contrast, the $x$ line-of-sight shows little variability in regards to turbulence or magnetization. We also demonstrate that beam convolution narrows the $p$ distributions (bottom panel of Figure \ref{fig:pPDF}), and therefore exacerbates the degree to which the $y$ and $z$ lines-of-sight disagree with the BLASTPol observations. Finally, we demonstrate the existence of linearity in the depolarized tail of the $p$ distribution, which is a signature for power-law behaviour in the PDF. }

\item{Similarly, our examination of the dispersion in polarization angles $S$ (Section \ref{section:S}) demonstrates that the mean angle dispersion in Vela C is rather high, and is again remarkably consistent with the $x$ line-of-sight (Figure \ref{fig:SPDF}). We find that the $y$ and $z$ lines-of-sight produce $S$ distributions with relatively very low mean angle dispersions, which can be affected by the level of turbulence and magnetization: higher turbulence/lower magnetization (Model A vs. Model B) drives the mean angle dispersion higher. On the other hand, the $x$ line-of-sight shows exceptionally little dependence on turbulence or magnetization. Interestingly, we find that the width of the distributions varies little between all lines-of-sight of both simulations, being substantially affected only by beam convolution (bottom panel of Figure \ref{fig:SPDF}), which tends to improve agreement with the BLASTPol observations. Finally, we note that near the value $S = \pi/\sqrt{12}$ (the value of $S$ where a collection of random vectors converges to) exists filamentary features in $S$, similar to structures that have been observed by Planck \citep{PXIX} and BLASTPol \citep{FBP}.}

\item{We further confirm remarkable consistency between the $x$ line-of-sight and the BLASTPol observations of Vela C in our examination of the joint correlation between polarization fraction and dispersion in polarization angles in Section \ref{section:Sp} (Figure \ref{fig:SP}). The power-law indices and correlation coefficients (Table \ref{tab:SpPCA}) match those of BLASTPol very well for both simulations. This correlation is still strong for the $y$ and $z$ lines-of-sight, but the power-law indices are much shallower. In these lines-of-sight it appears that the steepness of the slope is affected by higher turbulence/lower magnetization, with steeper slopes found in Model A. The power-law indices in the $x$ line-of-sight are not much affected by the level of turbulence and magnetization, consistent with our results for the polarization fraction and dispersion in polarization angles alone. We also note that beam convolution has little effect on the joint correlations (Figure \ref{fig:SP}, bottom panel). Altogether, we present strong evidence for relatively high apparent magnetic disorder in Vela C, but we note that this apparent magnetic disorder may not be a true proxy for intrinsic magnetic disorder arising from weak magnetization or strong turbulence. There is a degeneracy between this type of magnetic disorder and disorder that arises due to large inclination of the mean magnetic field with respect to the plane-of-sky, leading to the absence of an ordered plane-of-sky magnetic field.}

\item{Examining the joint correlations involving column density (Section \ref{section:N}), we find that none of the lines-of-sight particularly match the BLASTPol observations. The polarization fraction vs. column density correlations are generally weakly correlated (Figure \ref{fig:IP}). In the $y$ and $z$ lines-of-sight, the power-law index is much shallower than measured in BLASTPol, and for the $x$ line-of-sight, the magnitude of the power-law index is very large, but this has little weight given the weakness of the correlation as measured by the correlation coefficients (Table \ref{tab:IpPCA}). On the other hand, where BLASTPol saw little correlation in the dispersion in polarization angles - column density joint correlation, in some cases the simulations display a moderate positive correlation (e.g., the $z$ line-of-sight in Model A in Figure \ref{fig:IS} and Table \ref{tab:ISPCA}). We argue that we are unable to match the BLASTPol observations because we assumed homogeneous grain alignment, and that while there are some purely MHD contributions to these relationships, they are more likely dominated by the effects of grain alignment physics. We will explore this in a future publication.}

\item{By examining intermediate lines-of-sight between the $x$ line-of-sight and the $z$ lines-of-sight (Section \ref{section:inc}), we establish that agreement with the BLASTPol data may be found within a range of inclination angles, not just the pure $x$ line-of-sight (Figure \ref{fig:inclination}). Model A is consistent with BLASTPol at a wider range of inclination angles than Model B; Model A is consistent with at least inclinations greater than 75$^{\circ}$ measured from the $z$ to the $x$ line-of-sight. As inclination increases toward the $z$ lines-of-sight, disagreement is worsened. We argue that the inclination angle may be interpreted as a mixing angle rather than a physical inclination, indicating the relative degree to which the mean magnetic field is in the plane-of-sky, and that agreement is found when enough of the ordering influence of the mean magnetic field is suppressed by projection effects. We also find little evidence that inclination can achieve agreement between the simulations and BLASTPol joint distributions involving column density (Figure \ref{fig:inclination2}), adding further evidence that MHD structure alone cannot account for the observed correlations with column density.}
\end{enumerate}

To summarize, our comparisons of colliding flow simulations to the BLASTPol observations suggest that Vela C has a high degree of apparent magnetic field disorder, as supported by studies of the polarization fraction and dispersion in polarization angles. Whether this magnetic disorder is merely apparent (due to fortuitous alignment of the magnetic field relative to the line-of-sight rather than significant intrinsic magnetic field disorder) cannot be determined from our studies alone. We also find that the correlations involving column density cannot be explained by MHD structure alone; future work is needed modelling the effects of heterogeneous grain alignment.  

\section*{Acknowledgements}

The authors would like to thank Giles Novak, Juan-Diego Soler, and Blakesley Burkhart for helpful discussion and comments, and are grateful to the BLASTPol collaboration for making their data available. P. K. K. is supported in part through ALMA Student Observing Support provided by the National Radio Astronomy Observatory and a SAO predoctoral fellowship, and acknowledges additional support from the Jefferson Scholars Foundation and the Virginia Space Grant Consortium through graduate fellowships. P. K. K. is grateful to Qizhou Zhang for serving as his host at the Harvard-Smithsonian Center for Astrophysics. L. M. F. is a Jansky Fellow of NRAO. NRAO is a facility of the National Science Foundation (NSF operated under cooperative agreement by Associated Universities, Inc). C.-Y. C. thanks the support from Virginia Institute of Theoretical Astronomy (VITA) at the University of Virginia through the VITA Postdoctoral Prize Fellowship. Z.-Y. L. is supported in part by NASA 14AB38G and NSF AST1313083 and 1716259.

\bibliographystyle{mnras}
\bibliography{references}

\appendix

\bsp	
\label{lastpage}
\end{document}